\documentclass[11pt,fleqn,reqno]{article}
\pdfoutput=1
\usepackage[dvips]{graphicx}
\usepackage{epsfig}
\usepackage{psfig}
\usepackage{amsmath}
\usepackage{amsfonts}   
\usepackage{amssymb}    
\usepackage{bm}
\usepackage{subfigure}
\usepackage{color}
\usepackage{citesort}
\def\kl#1#2{{\mbox{\tiny \it #1#2}}}
\def\k#1{{\mbox{\tiny \it #1}}}
\newcommand{\order}{{\cal O}}
\def\b#1{{\bf #1}}

\begin{document}

{\large
\noindent 

\noindent{\large \bf Title:} A balanced memory network\\

\noindent{\bf Authors:} \\
\noindent Yasser Roudi,\\
Gatsby Computational Neuroscience Unit, UCL, UK\\

\noindent Peter E. Latham,\\
Gatsby Computational Neuroscience Unit, UCL, UK\\


\noindent{Acknowledgments:
YR and PEL were supported by the Gatsby Charitable Foundation and
National Institute of Mental Health Grant R01
MH62447. We thank Alfonso Renart and anonymous reviewers for very constructive comments.}\\
}

\newpage

\begin{abstract}

A fundamental problem in neuroscience is understanding how working
memory -- the ability to store information at intermediate timescales,
like 10s of seconds -- is implemented in realistic neuronal networks.
The most likely candidate mechanism is the attractor network, and a
great deal of effort has gone toward investigating it theoretically.
Yet, despite almost a quarter century of intense work, attractor
networks are not fully understood. In particular, there are still two
unanswered questions. First, how is it that attractor networks exhibit
irregular firing, as is observed experimentally during working memory
tasks? And second, how many memories can be stored under biologically
realistic conditions? Here we answer both questions by studying an
attractor neural network in which inhibition and excitation balance
each other. Using mean field analysis, we derive a three-variable
description of attractor networks. From this description it follows
that irregular firing can exist only if the number of neurons involved
in a memory is large. The same mean field analysis also shows that the
number of memories that can be stored in a network scales with the
number of excitatory connections, a result that has been suggested for
simple models but never shown for realistic ones. Both of these predictions
are verified using simulations with large
networks of spiking neurons. \end{abstract}

A critical component of any cognitive system is working memory -- a
mechanism for storing information about past events, and for accessing
that information at later times. Without such a mechanism even simple
tasks, like deciding whether to wear a heavy jacket or a light sweater
after hearing the weather report, would be impossible. Although it is
not known exactly how storage and retrieval of information is
implemented in neural systems, a very natural way is through attractor
networks. In such networks, transient events in the world trigger
stable patterns of activity in the brain, so by looking at the pattern
of activity at the current time, other areas in the brain can know
something about what happened in the past.

There is now considerable experimental evidence for attractor networks
in areas such as inferior temporal cortex \cite{Miy+88,Sak+91,Miy88},
prefrontal cortex \cite{Fus+71,Kub+71,Fun+89,Mil+96,Rai+98,Rao97}, and
hippocampus \cite{Leu05,Wil05}. And from a theoretical standpoint, it
is well understood how attractor networks could be implemented in
neuronal networks, at least in principle. Essentially, all that is
needed is an increase in the connection strength among subpopulations
of neurons. If the increase is sufficiently large, then each
sub-population can be active without input, and thus ``remember''
events that happened in the past.

While the basic theory of attractor networks has been known for some
time \cite{Hop82,Ami89,Ami97}, moving past the ``in principle''
qualifier, and understanding how attractors could be implemented in
realistic, spiking networks, has been difficult. This is
because the original Hopfield model violated several important
principles: neurons did not obey Dale's law; when a memory was
activated neurons fired near saturation, much higher than is observed
experimentally in working memory tasks \cite{Miy+88,Nakamura95}; and
there was no null background state -- no state in which all neurons
fired at low rates.

Most of these problems have been solved. The first, that Dale's law
was violated, was solved by ``clipping'' synaptic weights; that is, by
using the Hopfield prescription \cite{Hop82}, assigning neurons to be
either excitatory or inhibitory, and then setting any weights of the
wrong sign to zero \cite{Somp86,Burkitt96}. The second, building
a Hopfield type network with low firing rate, was solved by adding
appropriate inhibition
\cite{AmitTreves89,TrevesAmit89,Rubin89,Golomb90,Bru00,Lat04}
(importantly, this was a nontrivial fix; for discussion, see
\cite{Lat04}). The third problem, no null background, was solved
either by making the units sufficiently stochastic
\cite{AmitTreves89,TrevesAmit89,Rubin89,Golomb90} or adding external
input \cite{Rubin89,Golomb90,Ami97,Bru00,Lat04}.

In spite of these advancements, there are still two fundamental open
questions. One is: how can we understand the highly irregular firing
that is observed experimentally in working memory tasks \cite{Comp03}?
Answering this question is important because irregular firing is
thought to play a critical role both in how fast computations are
carried out \cite{Van96} and in the ability of networks to perform
statistical inference \cite{Ma06}. Answering it is hard, though,
because, as pointed out in \cite{Van05}, with naive scaling the net
synaptic drive to the foreground neurons (the neurons that fire at
elevated rate during memory) is proportional to the number of
connections per neuron. Consequently, because of the high connectivity
observed in cortex, the mean synaptic drive is much larger than the
fluctuations, which implies that the foreground neurons should fire
regularly. Moreover, as pointed out by Renart et al., \cite{Renart06},
even for models that move beyond the naive scaling and produce
irregularly firing neurons,
the foreground
neuron still tend to fire more regularly than the background neurons,
something that is inconsistent with experiments \cite{Comp03}.

Several studies have attempted to get around this problem,
either directly or indirectly
\cite{Bru00,BrunelWang01,Van05,Renart06}. Most of them, however did
not investigate the scaling of the network parameters with its size
(i.e., with the number of neurons and connections). So, although
parameters were found which led to irregular activity, it was not
clear how those parameters should scale as the size of the network
increased to realistic values.
In the two that did investigate scaling
\cite{Van05,Renart06},
irregular firing was possible only if a small fraction of
neurons was involved in each memory; i.e., only if
the coding level was
very small. Although there have been no direct measurements of
the coding level during persistent activity, at least to our
knowledge, experiments in superior temporal sulcus \cite{Rol95}
suggest that it is much larger than the one used these models.
We should point out, though, that the model of Renart et al.\
\cite{Renart06} is the only one in which
the foreground neurons are at least as regular as
the background neurons.


The second open question is: what is the storage capacity of realistic
attractor networks? That is, how many different memories can be stored
in a single network? Answering this is critical for understanding the
highly flexible and seemingly unbounded memory capacity observed in
animals. For simple, albeit unrealistic, models the answer is known:
as shown in the seminal work of Amit, Gutfreund and Sompolinsky
\cite{Ami+85}, the number of memories that can be stored
in a classical Hopfield network \cite{Hop82} is about 0.14
times the number of neurons. For slightly more realistic networks the
answer is also known \cite{TrevesAmit89,Golomb90,Somp86,Der+87,Tso+88,Tre+91,Tre91a,Curti04,Buh89,Tre90,Van05}.
However, even these more realistic studies lacked
biological plausibility in at least one way:
connectivity was all-all rather than sparse \cite{TrevesAmit89,Golomb90,Tso+88,Tre90},
the neurons were binary (either on or off, with nothing in between) \cite{TrevesAmit89,Somp86,Golomb90,Tso+88,Buh89,Der+87},
there was no null background \cite{Tre90,Tre91a,Somp86,Buh89,Tso+88,Der+87},
the firing rate in the foreground state was higher than is observed
experimentally \cite{Curti04,Somp86,Buh89,Tso+88,Der+87,Van05},
or the coding level was very small \cite{Curti04,Van05}.

Here we answer both questions: we show, for realistic networks of
spiking neurons, how irregular firing can be achieved, and we compute
the storage capacity. Our analysis uses relatively standard mean field
techniques, and requires only one assumption: neurons in the network
fire asynchronously. Given this assumption, we first show that neurons
fire irregularly only if the coding level is above some threshold,
although a feature of our model is that the foreground neurons
are slightly more regular than the background neurons.
We then show
that the maximum number of memories in our network -- the capacity --
is proportional to the number of connections per neuron, a result that
is consistent with the simplified models discussed above. These
predictions are verified with simulations of biologically plausible
networks of spiking neurons.

\section{Model} 
\label{Model}

To address analytically the issues of irregularity and storage
capacity in attractor networks, we consider a model in which neurons
are described by their firing rates. Although firing rate models
typically provide a fairly accurate description of network behaviour
when the neurons are firing asynchronously \cite{Tre93,Shriki03}, they
do not capture all features of realistic networks. Therefore, we
verify all of our predictions with large-scale simulations of spiking
neurons.

Our network consists of two populations, one excitatory and one
inhibitory, with $N_E$ neurons in the former and $N_I$ in the latter.
(In general we use $E$ for excitation and $I$ for inhibition.) We
represent the firing rate of the $i^{\rm th}$ neuron in pool $Q(=E,I)$
by $\nu_{Qi}$. As we show in Appendix I, and discuss below,
the time evolution equations for the firing rates
are given by

\begin{subequations}
\begin{align}
\tau_\k E \, \frac{d\nu_{\k Ei}}{dt} & = F_E\left(h_{Ei}\right)-\nu_{\k Ei}
\\
\tau_\k I \, \frac{d\nu_{\k Ii}}{dt} & = F_I\left(h_{Ii}\right)-\nu_{\k Ii}
\, ,
\end{align}
\label{dynamic_fr}
\end{subequations}

\noindent
where $\tau_E$ and $\tau_I$ are the excitatory and inhibitory time
constants, $h_{Qi}$ is the synaptic input to the $i^{\rm th}$ neuron
in pool $Q$, and $F_Q(h)$ is a function that tells us the steady state
firing rate of a neuron receiving synaptic input $h$. This function,
which has a relatively stereotyped quasi-sigmoidal shape,
can be determined analytically (or semi-analytically) for specific
noise models \cite{Tuck88,Fourcaud02,Bru03}, and 
numerically for more realistic models \cite{Shriki03}.
The synaptic drive, $h_{Qi}$, is related to the activity of the
presynaptic neurons via
\begin{equation}
h_{Qi}=\sum^{N_E}_{j=1} \tilde{J}^{QE}_{ij}
\nu_{\k Ej}+\sum^{N_I}_{j=1} \tilde{J}^{QI}_{ij}\nu_{\k Ij}+ \tilde{h}_{Qex}
\, ,
\label{loc-field}
\end{equation}

\noindent
where $ \tilde{J}^{QR}_{ij}$ is the synaptic weight from the $j^{\rm th}$
neuron in pool $R$ to the $i^{\rm th}$ neuron in pool $Q$ and
$\tilde{h}_{Qex}$ is the external, purely excitatory,
input to neurons in pool $Q$. Finally,
the steady-state firing rate of each neuron is determined by setting
$d \nu_{\k Ei}/dt$ and $d \nu_{\k Ii}/dt$ to zero, yielding the
equation
\begin{equation}
\nu_{Qi} = F_Q(h_{Qi})
\label{steady_state}
\, .
\end{equation}

The bulk of our analysis focuses on solving Eq.\ (\ref{steady_state});
we use the dynamics, Eq.\ (\ref{dynamic_fr}), only when
investigating stability. Our goal is to determine the conditions that
support retrieval states -- states such that subpopulations of neurons
have elevated firing rates.

Since the gain functions, $F_Q(h)$, that we use in Eq.\
(\ref{dynamic_fr}) play such a central role in our analysis, we
briefly justify them here; for additional details, see Appendix I.
These gain functions come from an average over the fast
temporal fluctuations of the synaptic input -- basically, filtered
spikes. Calculating the temporal fluctuations self-consistently
is a hard problem \cite{Hertz04} but, fortunately, it's not a problem we
have to solve. As we show in Appendix I, in
the limit that each neuron receives a large number of connections, the
temporal fluctuations experienced by all the excitatory neurons have
the same statistics, as do the temporal fluctuations experienced by
all the inhibitory neurons.
Thus, we can use a single function,
$F_E(h)$, for the
excitatory neurons, and another function,
$F_I(h)$, for the inhibitory ones.
Of course, we won't be able to calculate the shape of
$F_Q$ without knowing the structure of the temporal fluctuations.
However, as we show below,
the precise shapes of the gain functions don't play a strong role in
our analysis.

\subsection{Connectivity} 

The main determinant of network behaviour, at least in this model, is
the set of connection strengths, the $ \tilde{J}^{QR}_{ij}$ (as we
will see, single neuron properties affect the quantitative details,
but do not play a strong role in the qualitative behaviour). To choose
connection strengths that will lead to attractors, we build on the
model proposed by Hopfield over two decades ago \cite{Hop82}. In that
model, random patterns are stored via a Hebbian learning rule, so
connection strengths among neurons have the form
\begin{equation}
A_{ij}=\tilde{\beta} \sum^p_{\mu=1} \xi^{\mu}_{i} (\xi^{\mu}_{j}-a)
\, ,
\label{A-ij}
\end{equation}

\noindent
where $A_{ij}$ is the strength of the connection from
neuron $j$ to neuron $i$,
$\xi^{\mu}_{i}=1$ if neuron $i$ participates in
pattern $\mu$ and $\xi^{\mu}_{i}=0$ otherwise, $\tilde{\beta}$ is a constant
that determines the memory strength, and $p$ is the number of patterns.
For each neuron, the probability of participating in a given pattern,
$\mu$, is equal to the coding level, which we denote $a$. Thus,

\begin{equation}
\xi^{\mu}_{i} = \left\{ \begin{array}{llr}
1 &
\ \ {\rm with \ probability \ } a
\\
0 & \ \ {\rm with \ probability \ } 1-a
\, .
\end{array} \right.
\label{prob_a}
\end{equation}

\bigskip

\noindent
With this definition, the
term $(\xi^{\mu}_{j}-a)$ in Eq.\ (\ref{A-ij}) ensures that, on average,
$\sum_j A_{ij}$ is zero. Thus, the learning
rule does not change the total synaptic weight onto a neuron, a form
of postsynaptic normalisation that has been observed experimentally
in cultured networks \cite{turrigiano98,desai99}.

While Eq.\ (\ref{A-ij}) produces a network that exhibits attractors,
it is inconsistent with biology in at least two important ways. First,
the neurons can exhibit both excitatory and inhibitory
connections (for fixed presynaptic neuron $j$, $A_{ij}$ can be
positive for some postsynaptic targets $i$ and negative for others),
which violates Dale's law. Second, connectivity is all to all,
which is inconsistent with the sparse connectivity seen in cortex
\cite{Bra91}. Both can be fixed by introducing
sparse, random background connectivity among excitatory and inhibitory
neurons, and adding a threshold so that neurons are either
excitatory or inhibitory, but not both. This yields a set of
connection strengths of the form

\begin{subequations}
\begin{align}
\tilde{J}^{EE}_{ij} & = c^{EE}_{ij} [\tilde{J}_{EE}+A_{ij}]^+
\\
\tilde{J}^{IE}_{ij} & = c^{IE}_{ij} \tilde{J}_{IE}
\\
\tilde{J}^{EI}_{ij} & = c^{EI}_{ij} \tilde{J}_{EI}
\\
\tilde{J}^{II}_{ij} & = c^{II}_{ij} \tilde{J}_{II}
\, ,
\end{align}
\label{syn-weights}
\end{subequations}


\noindent
where the $\tilde{J}_{QR}$ set the background connection strengths (with, of
course, $\tilde{J}_{EE}$ and $\tilde{J}_{IE}$ positive and
$\tilde{J}_{EI}$ and $\tilde{J}_{II}$ negative), $[ \cdot ]^+$ is
the threshold-linear operator ($[ x ]^+ = x$ if $x > 0$ and 0
otherwise), and
$c_{ij}^{QR}$ is the probability that neuron $j$ of type $R$
connects to neuron $i$ of type $Q$. We assume that the connection
probability is independent of type, so

\begin{equation}
c^{QR}_{ij}= \left\{ \begin{array}{llr}
1 &
\ \ {\rm with \ probability \ } c
\\
0 & \ \ {\rm with \ probability \ } 1-c
\, .
\end{array} \right.
\label{prob_c}
\end{equation}

\bigskip
\noindent With this connectivity matrix, every neuron in the network projects
to, on average, $K_E$ excitatory and $K_I$ inhibitory neurons, and
every neuron receives, on average, $K_E$ excitatory and $K_I$
inhibitory connections, where

\begin{subequations}
\begin{align}
K_E\equiv c N_E
\\
K_I\equiv c N_I.
\end{align}
\label{keki}
\end{subequations}

\noindent
The probability of connection, $c$, is assumed to be much smaller than
1, leading to a sparsely connected network \cite{Bra91}, and it is
independent of the size of the network unless otherwise stated. While
we could have made the connectivity scheme more general by letting
the connection probability between neurons depend on their type
and/or by letting the nonzero $c_{ij}^{QR}$ in Eq.\
(\ref{prob_c}) have some variability, this would merely add
complexity without changing any of our conclusions.

Although we are including the threshold-linear operator in Eq.\
(\ref{syn-weights}) (and also in the simulations) we neglect it in the
forthcoming theoretical analysis. This is because $A_{ij}$ tends to be
small: Its mean is zero and, as we discuss in section
\ref{Str_Cap_Sec} and Appendix II, its variance is
$\order(p/K_\k E)$. Thus, as long as $p$ is sufficiently small compared
to $K$, the threshold-linear operator can be neglected. For our model,
we find that $p/K$ is at most about 0.01,
which means that the threshold-linear operator is unlikely to
have much effect. Importantly, even if $p/K$ were large,
the scaling relation that we derive for
storage capacity, i.e. $p_{\max} \sim K$, would still be correct; the
only effect would be a slight modification to the precise value of
$p_{\max}/K$ \cite{Somp86}.

\section{Network equilibria}

As discussed above, much of our focus in this paper is on solving
Eq.\ (\ref{steady_state}).
For even moderate size networks, this corresponds
to solving thousands of coupled, highly nonlinear equations, and for large
networks that number can run into the millions. We do not, therefore,
try to find a particular solution to this equation, but instead look
for a statistical description -- a description in terms of probability
distributions over excitatory and inhibitory firing rates. The main
tool we use is self-consistent signal-to-noise analysis
\cite{Shi+92,Shi+93}. The idea behind this analysis is to treat the
synaptic input ($h_{Ei}$ and $h_{Ii}$ in Eq.\ (\ref{steady_state})) as
Gaussian random variables. Solving Eq.\ (\ref{steady_state}) then
reduces to finding, self-consistently, their means and variances.

Because
$h_{Ei}$ and $h_{Ii}$ consist of $2K$ (very weakly) correlated
terms, where
\begin{equation}
K \equiv \frac{K_E + K_I}{2}\nonumber
\, ,
\end{equation}

\noindent
naive central limit arguments tell us that the standard
deviations of these quantities should be smaller than their means by a
factor of $K^{1/2}$. It would seem, then, that in the kinds of high
connectivity networks found in the brain, where $K$ is on the order of
5,000-10,000, neuron to neuron fluctuations in firing rate would be
small, on the order of $K^{-1/2}$. By the same reasoning, {\em
temporal} fluctuations in the firing rates would also be small, again
on the order of $K^{-1/2}$. Neither of these, however, are observed in
biological networks: there are large fluctuations in firing rate both
across neurons and over time \cite{Comp03,Noda70,Burn76,Soft93,Hol96}.

To resolve this apparent contradiction, one need only notice that
$h_{Ei}$ and $h_{Ii}$ consist of both positive and
negative terms (the first and third terms in Eq.\ (\ref{loc-field}) are
positive; the second is negative).
If these terms approximately cancel -- to within
$\order(K^{-1/2})$ -- then both the mean and standard deviation of the
synaptic drive will be on the same order, and network irregularity
will be restored. As showed by van Vreeswijk and Sompolinsky in a
groundbreaking set of papers \cite{Van96,Van98}, under fairly mild
conditions this cancellation
occurs {\em automatically}, thus placing networks very naturally in
what they called the {\em balanced} regime. In this regime,
fluctuations across both neurons and time are large. Whether
networks in the brain really operate in the balanced regime is not 
completely clear, although recent experimental evidence has come down 
strongly in favour of this hypothesis \cite{You93,Bil06}.

While the work of van Vreeswijk and Sompolinsky was extremely
important in shaping our understanding of realistic recurrent
networks, their focus was primarily on random connectivity. The
situation, however, is more complicated in attractor networks. That's
because these networks consist of three classes of neurons rather than
two: background excitatory neurons and background inhibitory neurons,
as found in randomly connected networks, but also foreground
excitatory neurons.
Our goal in the next several sections is to understand how all three
classes can be balanced, and thus fire irregularly.


\subsection{Strong synapses and the balanced condition}

A reasonable constraint to place on our theoretical framework is that,
in the large $K$ limit, our results should be independent of $K$.
This suggests that the synaptic strength, the $\tilde{J}_{QR}$ in
Eq.\ (\ref{syn-weights}), should scale as $K^{-1/2}$. With this
scaling, the mean value of the positive and negative terms in
$h_{Ei}$ and $h_{Ii}$ become $\order(K^{1/2})$, with cancellation these
terms are $\order(1)$, and the variance is also $\order(1)$. Thus,
if the gain functions, the $F_Q(h)$ in Eq.\ (\ref{steady_state}),
are also $\order(1)$, our results will be independent of the number of
connections. 
To make the $K^{-1/2}$ scaling
explicit, we define a new set of synaptic strengths
and external input, which we denote $J_{QR}$ and $h_{Qex}$,
respectively,

\begin{subequations}
\begin{align}
\tilde{J}_{QR} & = {K^{1/2} J_{QR} \over K_R}
\label{newJa}
\\
\tilde{h}_{Qex} & = K^{1/2} h_{Qex}
\end{align}
\label{newJ}
\end{subequations}

\noindent
where $J_{QR}$ and $h_{Qex}$ are both $\order(1)$ and, recall, $K_R= c
N_R$ (Eq.\ (\ref{keki})).

Equation (\ref{newJ}) tells us how to scale the background
connectivity, but it does not directly apply to the part of the
connection matrix associated with memories, $A_{ij}$. To determine
how $A_{ij}$ should scale,
we need only note that the mean contribution from
the memories should be $\order(1)$ -- sufficiently large to have an
effect,  but not so large as to overwhelm the background.
Consequently, $A_{ij}$ should scale as $1/K$ (see Appendix II for
details), which we can guarantee
by defining a new variable, $\beta$, via the relation

\begin{equation}
\tilde{\beta} \equiv \frac{\beta}{K_Ea(1-a)}
\label{beta-scaling}
\end{equation}

\bigskip

\noindent
where $\beta$ is $\order(1)$ and the factor
$a(1-a)$ is for convenience only.

\subsection{Mean field equations for the retrieval states}
\label{Meandriven}
Now that we have the ``correct'' scaling -- scaling that makes our
results independent of network size and ensures that the mean and
variance of the synaptic input are both $\order(1)$ -- we can apply
self-consistent signal-noise analysis to Eq.\ (\ref{steady_state}).
The first step is to
divide the excitatory and inhibitory synaptic currents ($h_{Ei}$ and
$h_{Ii}$) into two pieces: one that is nearly independent of
index, $i$ (the ``mean''),
and one that is a random variable with respect
to $i$ (the fluctuating piece). To do that,
we rewrite the synaptic current in terms of our new
variables, $J_{QR}$ and
$\beta$, rather than $\tilde{J}^{QR}_{ij}$ and $\tilde{\beta}$. Combining Eqs.\
(\ref{A-ij}), (\ref{syn-weights}), (\ref{newJ}) and
(\ref{beta-scaling}) with Eq.\ ({\ref{loc-field}), we have

\begin{subequations}
\begin{align}
h_{\k Ei}  & =
K^{1/2}
\left[
\sum_R {J_{ER} \over K_R} \sum_{j=1}^{N_R}
c^{ER}_{ij} \nu_\k {Rj}
+ h_{Eex}
\right]
+ {\beta \over K_E a (1-a)}
\sum_{\mu = 1}^p \sum_{j=1}^{N_E} c^{EE}_{ij}
\xi_i^\mu (\xi_j^\mu - a) \nu_\k {Ej}
\label{loc-field-separated.a}
\\
h_{\k Ii}  & =
K^{1/2}
\left[
\sum_R {J_{IR} \over K_R} \sum_{j=1}^{N_R}
c^{IR}_{ij} \nu_\k {Rj}
+ h_{Iex}
\right]
\, .
\label{loc-field-separated.b}
\end{align}
\label{loc-field-separated}
\end{subequations}

\noindent
Note that Eq.\ (\ref{loc-field-separated}) is identical to
Eq.\ (\ref{loc-field}); it is just expressed in different variables.

For the terms in brackets, the mean and fluctuating pieces are easy to
compute: the mean comes from replacing $c^{QR}_{ij}$ by its average, $c$,
and the fluctuating piece comes from replacing
$c^{QR}_{ij}$ by the residual, $c^{QR}_{ij} -c$.
For the second term in Eq.\ (\ref{loc-field-separated.a}), separating
the mean from the fluctuating piece is harder, as there is a
nontrivial dependence on $i$ associated with the $p$ memories.
Ultimately, however, we are interested in the
case in which only one memory is retrieved, so when computing the mean
we can consider only
one term in this sum on $\mu$; the other $p-1$ terms contribute only to the
fluctuations. Assuming, without loss of generality,
that the first memory is retrieved, averaging over the randomness 
associated with the sparse connectivity allows us to replace $c^{EE}_{ij}$ 
with $c$, and we find that the mean of the
last term in Eq.\ (\ref{loc-field-separated.a})
is proportional to $\xi_i^1$.

Putting all this together, we arrive at the eminently reasonable
result that the mean excitatory and inhibitory synaptic currents
are linear in the mean excitatory and inhibitory firing rates, and the
mean excitatory current has an extra, memory induced, dependence
proportional to $\xi_i^1$. Dropping the superscript ``1'' (a step
taken only to simplify the equations), we find that the synaptic
current may be written

\begin{subequations}
\begin{align}
h_{\k Ei} & = h_\k E
+ \xi_i \beta m
+ \delta \hat{h}_{\k Ei}
\label{mean_var.a}
\\
h_{\k Ii} & = h_\k I
+ \delta h_{\k Ii},
\end{align}
\label{mean_var}
\end{subequations}

\noindent
where $h_\k E$ and $h_I$ are the averages of the terms in
brackets on the right hand side of Eq.\ (\ref{loc-field-separated}), $\xi_i \beta m$ is the mean
contribution from the first memory, and $\delta \hat{h}_{\k Ei}$ and
$\delta h_{\k Ii}$ contain everything else. More
specifically, the terms in Eq.\ (\ref{mean_var}) are as follows.
First, $h_E$ and $h_I$ are given by

\begin{subequations}
\begin{align}
h_\k E & =
K^{1/2} (
J_\k {EE} \nu_\k E +
J_\k {EI} \nu_\k I +
h_{{\k E}ex})
\label{he_hi.a}
\\
h_\k I & =
K^{1/2} (
J_\k {IE} \nu_\k E +
J_\k {II} \nu_\k I +
h_{{\k I}ex}
)
\, ,
\label{he_hi.b}
\end{align}
\label{he_hi}
\end{subequations}

\noindent
where $\nu_\k E$ and $\nu_\k I$ are the firing rates averaged over the
excitatory and inhibitory populations, respectively,

\begin{subequations}
\begin{align}
\nu_\k E & \equiv
\frac{1}{N_E}\sum_i \nu_{\k Ei}
\\
\nu_\k I & \equiv
\frac{1}{N_I}\sum_i \nu_{\k Ii}
\, .
\end{align}
\label{ne_ni}
\end{subequations}

\noindent
Second, the overlap, $m$, which is proportional to the mean firing rate
of the foreground neurons relative to $\nu_\k E$, is given by

\begin{equation}
m \equiv 
{1 \over N_Ea(1-a)}
\sum_i (\xi_i-a) \nu_\k {Ei}
\, .
\label{overlap}
\end{equation}

\noindent
Expressions for
the fluctuating terms, $\delta \hat{h}_{\k Ei}$ and $\delta h_{\k Ii}$,
are given in Eqs.\ (\ref{var_back}) and (\ref{var_fore}). Because
these terms contain {\em everything} not contained in the
mean terms, Eq.\ (\ref{mean_var}) is exact.

The three quantities $\nu_\k E$, $\nu_\k I$ and $m$ are our main
order parameters. To determine their values self-consistently, we
express the firing rates, $\nu_\k {Ei}$ and $\nu_\k {Ii}$,
in terms of the synaptic currents using
Eq.\ (\ref{steady_state}), and insert those expressions back into
Eqs.\ (\ref{ne_ni}) and (\ref{overlap}); that leads to

\begin{subequations}
\begin{align}
\nu_\k E &= {1 \over N_E} \sum_i F_\k E( h_\k E + \xi_i \beta m +
\delta \hat{h}_{\k Ei})
\\
m &= {1 \over N_E a (1-a)} \sum_i
(\xi_i - a) F_\k E( h_\k E + \xi_i \beta m +
\delta \hat{h}_{\k Ei})
\\
\nu_\k I &= {1 \over N_I} \sum_i F_I( h_\k I + \delta h_{\k Ii})
\, .
\end{align}
\label{mf_sum}
\end{subequations}

To solve these equations, we use the fact that there are a large
number of neurons; this allows us to turn
the sum over $i$ into an integral over the probability distributions of
$\delta \hat{h}_E$ and $\delta h_I$, denoted $p(\delta \hat{h}_E)$ 
and $p(h_I)$, respectively. Replacing the sum by an integral in
Eq.\ (\ref{mf_sum}), and also averaging
over $\xi_i$, the mean field equations become

\begin{subequations}
\begin{align}
\nu_\k E &=
\int d \delta \hat{h}_\k E \ p(\delta \hat{h}_\k E) \,
\langle F_\k E( h_\k E + \xi \beta m + \delta \hat{h}_\k E) \rangle_\xi \nonumber
\\
m &=
\int d \delta \hat{h}_\k E \ p(\delta \hat{h}_\k E) \,
\left\langle
{\xi - a \over a(1-a)} F_\k E( h_\k E + \xi \beta m + \delta \hat{h}_\k E)
\right\rangle_\xi\nonumber
\\
\nu_\k I &=
\int d \delta h_\k I \ p(\delta h_\k I) \,
F_\k I( h_\k I + \delta h_\k I)\nonumber
\, ,
\end{align}
\end{subequations}

\noindent
where the subscript on the angle brackets
indicates an average with the
statistics given in Eq.\ (\ref{prob_a}).

Because both $\delta \hat{h}_\k E$ and $\delta h_\k I$ are Gaussian random
variables (see Appendix
I), these integrals are reasonably straightforward; what makes them
at all difficult is that the variance of $\delta \hat{h}_\k E$ and $\delta h_\k I$
must be found self-consistently. This results in two more equations,
for a total of five (see Eq.\ (\ref{MF})). This is still far fewer than
our original set of 1000s or more. And the situation gets even
better: it turns out that we
really only need to consider three, at least if all we want to do is gain
qualitative insight into how attractor networks function.
That's because the integrals
are simply Gaussian convolutions, so all
they do is smooth the gain functions. Using a bar to denote
the smoothed functions, and performing the average over $\xi$
(which is straightforward because it has simple 0/1 statistics; see
Eq.\ (\ref{prob_a})) we have

\begin{subequations}
\begin{align}
\nu_\k E &=
(1-a) \bar{F}_E( h_\k E) +
a \bar{F}_E( h_\k E + \beta m)
\label{mf_tilde_hE}\\
m &=
\bar{F}_E( h_\k E + \beta m) - \bar{F}_E( h_\k E)
\label{mf_tilde_m}
\\
\nu_\k I &=
\bar{F}_I( h_\k I).
\label{mf_tilde_hI}
\end{align}
\label{mf_tilde}
\end{subequations}

These equations -- which are identical in form to the ones derived in
\cite{Lat04} -- are oversimplified versions of the full
mean field equations. Basically, the bar over $F$
hides a dependence on two additional order parameters -- the second
moments of the excitatory and inhibitory firing rates -- which in turn
depend on our main order parameters, $\nu_\k E$, $\nu_\k I$,
and $m$. While these dependencies are important for making detailed
predictions, for an intuitive picture of what the mean field equations
mean they can be ignored. Consequently, in the next several sections
we focus on Eqs.\ (\ref{mf_tilde}a-c), which we refer to as the reduced mean
field equations. At the end of
the next section, we argue that, under very general conditions,
all the conclusions we draw based on
the reduced mean field equations apply to the full set
(which are given in Eq.\ (\ref{MF})).

\subsection{Reduced mean field equations in the infinite $K$ limit}
\label{red-MF}

When solving the reduced mean field equations we have a choice: we can
think of them as functions of $\nu_\k E$, $\nu_\k I$ and $m$, or as
functions of $h_\k E$, $h_\k I$ and $m$. Since $\nu_\k E$ and $\nu_\k I$ are
related to $h_\k E$ and $h_\k I$ via an invertible transformation --
Eq.\  (\ref{he_hi}) -- the two
prescriptions are identical. The latter, however, turns out to be more
convenient, especially in the infinite $K$ limit. To see why, we need
only solve Eq.\ (\ref{he_hi}) for the mean firing rates, which yields

\begin{subequations}
\begin{align}
\nu_\k E &= \nu_\k {E0} +
K^{-1/2} D^{-1} \left[ J_\k {II} h_\k E - J_\k {EI} h_\k I \right]
\\
\nu_\k I &= \nu_\k {I0} +
K^{-1/2} D^{-1} \left[ J_\k {EE} h_\k I - J_\k {IE} h_\k E \right]
\, 
\end{align}
\label{balanced_cond}
\end{subequations}

\noindent
where

\begin{subequations}
\begin{align}
\nu_\k {E0} & \equiv D^{-1} \left[ J_\k {EI} h_{\k Iex}-J_\k {II} h_{\k Eex} \right]
\\
\nu_\k {I0} & \equiv D^{-1} \left[ J_\k {IE} h_{\k Eex}-J_\k {EE} h_{\k Iex} \right]
\end{align}
\label{nu0}
\end{subequations}

\noindent
are the mean firing rates in the infinite $K$ limit and
\begin{equation}
D \equiv J_\k {EE} J_\k {II} - J_\k {EI} J_\k {IE}
\label{detJ}
\end{equation}

\noindent
is the determinant of the background connectivity matrix;
as shown in \cite{Van98} and Appendix III, $D$ must be positive for the background to be
stable. Since we are in the balanced regime,
$h_E$ and $h_I$ are $\order(1)$. Consequently, in the infinite $K$
limit, the mean excitatory and inhibitory firing rates are simply
given by $\nu_\k {E0}$ and $\nu_\k {I0}$, respectively, independent of
$h_E$ and $h_I$. Using this fact, the reduced mean field equations,
Eq.\ (\ref{mf_tilde}), become, in the $K \rightarrow \infty$ limit,

\begin{subequations}
\begin{align}
\nu_\k {E0} &=
(1-a) \bar{F}_E( h_\k E) +
a \bar{F}_E( h_\k E + \beta m)
\label{mf_tilde0_hE}\\
m &=
\bar{F}_E( h_\k E + \beta m) - \bar{F}_E( h_\k E)
\label{mf_tilde0_m}
\\
\nu_\k {I0} &=
\bar{F}_I( h_\k I)
\label{mf_tilde0_hI}\, .
\end{align}
\label{mf_tilde0}
\end{subequations}

An important feature of these equations is that $h_\k I$ decouples from
$h_\k E$ and $m$. This greatly simplifies the analysis, since it means we
can find the equilibrium value of
$h_\k I$ simply by inverting $\bar{F}_I$. 

Our approach to finding the equilibrium values of $h_E$ and $m$
is a graphical
one: we plot, in $h_E$-$m$ space, the two curves that correspond to the
solutions to Eqs.\ (\ref{mf_tilde0_hE}) and (\ref{mf_tilde0_m}) -- the $h_E$ and $m$
nullclines, respectively -- and look for their
intersections. The goal is to determine the conditions under which
there are multiple intersections, with at least one of them
corresponding to an equilibrium with $m > 0$, and thus to a retrieval
state.

To be as general as possible, we make only two assumptions:
$\bar{F}_E(h)$ is monotonic increasing, and it is
quasi-sigmoidal, where we use ``quasi-sigmoidal'' to mean
convex ($\bar{F}''_E(h) > 0$)
for small $h$ and concave ($\bar{F}''_E(h) < 0$) for large $h$.
(Note that $\bar{F}_E(h)$ need not saturate.)
This immediately tells us something about the
shape of the $h_E$-nullcline: since the
right hand side of Eq.\ (\ref{mf_tilde0_hE}) is an increasing function
of both $h_E$ and $m$, its solution, $h_E(m)$, must
have negative slope (i.e., $dh_E/dm < 0$ along the $h_E$ nullcline).
Typical plots of the $h_E$-nullcline are shown in
Fig.\ 1a for two values of the coding level, $a$. 
Note that the nullcline curves upward
in this plot, a consequence of the fact that we use
-$h_E$ rather than $h_E$ on the $y$-axis. 

\begin{figure}

\includegraphics[height=9cm,width=12cm]{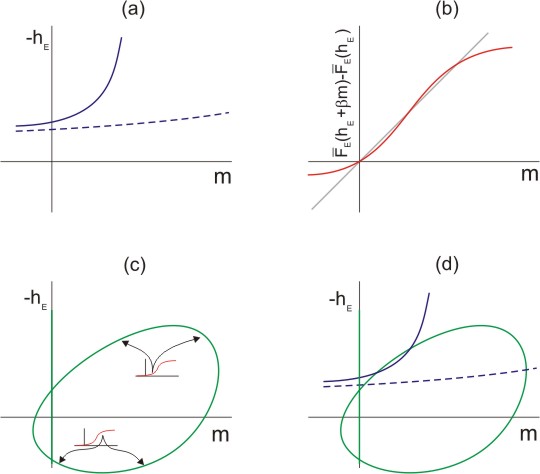}%
\caption{Generic shapes of the nullclines. Note that these are
``cartoons,'' and thus do not apply to any particular model; for
nullclines derived from a specific model, see Fig.\ 2. {\bf (a)} $h_\k
E$-nullcline versus $m$ for two different value of $a$ ($a$ is small
for the dashed curve and large for the solid curve). Note that we use
-$h_\k E$ on the $y$-axis, so the upward curvature indicates that the
total synaptic drive to a cell decreases with $m$. {\bf (b)} Right
hand side of Eq.\ (22b) versus $m$ with $h_\k E$ fixed.
The intersections with the $45^\circ$ line correspond to points on the
$m$-nullcline. {\bf (c)} The $m$-nullcline. The precise shape isn't so
important; what is important is that the part of the nullcline not on
the $m=0$ axis has the topology of a circle. Insets indicate the
portion of $F_\k E(h_\k E)$ that contributes to the $m$-nullcline; see
text. {\bf (d)} The $m$- and $h_\k E$-nullclines on the same plot. The
intersections correspond to network equilibria. There are three
equilibria: one at $m=0$, corresponding to the background state, and
two at $m > 0$, corresponding to potential retrieval states. The one
at $m=0$ and the one at large $m$ are stable; the intermediate one is
not. Consequently, only the large $m$ equilibrium is observed during
retrieval. Note that when the coding level, $a$,
is small (dashed blue line), the
retrieval state occurs at large $m$, and thus high firing rate. Only
when $a$ is large (solid blue line) is it possible to have low firing
rate during retrieval.
}
\label{Fig1}
\end{figure}

To find the $m$-nullcline -- the set of points in $h_E$-$m$ space that
satisfy Eq.\ (\ref{mf_tilde0_m}) -- we proceed in two stages. First,
we plot the right hand side of Eq.\ (\ref{mf_tilde0_m}) versus $m$ and look for
intersections with the $45^{\circ}$ line; these intersections
correspond to points on the $m$-nullcline. Second, we vary $h_E$ and
sweep out a curve in $h_E$-$m$ space; this curve is the full
$m$-nullcline. A typical plot versus $m$ with $h_E$ fixed
is shown in in Fig.\ 1b.
There are three intersections with the $45^{\circ}$
line, which means that the $m$-nullcline consists of three
points at this particular value of $h_E$: one with
$m=0$ and two with $m > 0$. To find out how these three points move as
we vary $h_E$, we compute $dm(h_E)/dh_E$
where the derivative is taken along the $m$-nullcline; using
Eq.\ (\ref{mf_tilde0_m}), this is given by

\begin{equation}
{dm(h_E) \over dh_E} =
{
\partial [\bar{F}_E( h_\k E + \beta m) - \bar{F}_E( h_\k E)]/\partial h_E
\over 1 -
\partial [\bar{F}_E( h_\k E + \beta m) - \bar{F}_E( h_\k E)]/\partial m
}
\, .
\label{dhEdm}
\end{equation}

\bigskip

We are primarily interested in the sign of $dm/dh_E$, which can be
found by examining the signs of the numerator and denominator
separately. For
the denominator, note that the derivative of the
term in square brackets is the slope of the curve in Fig. 1b.
Consequently, the
denominator is negative for the intermediate intersection (where the
slope is greater than 1) and positive for the upper intersection
(where the slope is less than 1). The sign of the numerator depends
primarily on the size of $h_E$. If $h_E$ is small, so that both
$\bar{F}_E(h_E+\beta m)$ and $\bar{F}_E(h_E)$ lie on the convex part
of the sigmoid, then the numerator is positive. If, on the other
hand, $h_E$ is large, so that
$\bar{F}_E(h_E+\beta m)$ and $\bar{F}_E(h_E)$ lie on the concave part,
then it is negative (see insets in Fig.\ 1c).

This gives us the following picture: when $h_E$ is small, so that the
numerator in Eq.\ (\ref{dhEdm}) is positive, decreasing $h_E$
causes the two intersections in Fig.\ 1b to move closer, and
eventually annihilate. When $h_E$ is large, on the other hand, so that
the numerator is negative, increasing, rather than decreasing, $h_E$
causes the intersections to move closer, and eventually annihilate,
this time for sufficiently large $h_E$. Filling in the points away
from the extrema, we see that the $m$-nullcline is topologically
equivalent to a circle (Fig.\ 1c). Finally we note that the
line $m=0$ is also part of the nullcline, as can easily be seen from
Eq.\ (\ref{mf_tilde0_m}); this line is also included in Fig.\ 1c.

In Fig.\ 1d we combine the $h_E$-nullclines from Fig.\ 1a and the
$m$-nullcline form Fig.\ 1c. Clearly there is always an equilibrium at
$m=0$, corresponding to no active memories; i.e., corresponding to a
null background. There are also two
equilibria at $m>0$, corresponding to active memories. In Appendix III
we show that the one at larger $m$ is stable. Importantly, this
equilibrium can occur at small $m$, and thus low firing rate,
something we will see more quantitatively in the next section, where
we consider a specific example. Although not shown in Fig.\ 1, the
$m$-nullcline can shift far enough up so that $m$ can be negative at
equilibrium. When this happens, $m=0$ becomes unstable, which in turn
implies that the background becomes unstable. We see this in the
simulations: when $\beta$ becomes too large, memories are
spontaneously activated.

We can now see the critical role played by the coding level, $a$. In
the limit $a \rightarrow 0$, the right hand side of
Eq.\ (\ref{mf_tilde0_hE}) becomes almost
independent of $m$. This makes the $h_E$-nullcline almost
horizontal (dashed line in Fig.\ 1d),
so the only stable retrieval state occurs at large $m$, and
thus high firing rate (the peak of the $m$-nullcline typically occurs
near the maximum firing rate of the neurons, about 100 Hz; see next
section). If, on the other hand, $a$ is reasonably large,
then the $h_E$-nullcline can curve up and intersect the
$m$-nullcine to the left of its highest point (solid blue line in
Fig.\ 1d). As just discussed, this intersection corresponds to the
intermediate intersection in Fig.\ 1b, which means it corresponds to
low firing rate, and thus a biologically realistic retrieval state.

We end this section by discussing the conditions under which the
nullclines in Fig.\ 1d, which were derived from Eq.\ (\ref{mf_tilde}),
are the same as the nullclines for the full mean field equations, Eq.\
(\ref{MF}). The primary effect of the full set of equations is to
couple $h_I$ to $h_E$ and $m$. One could, however, solve for $h_I$ in
terms of $h_E$ and $m$, insert that solution into the equations for
$h_E$ and $m$, and derive a new coupled set of equations that again
involve only $h_E$ and $m$.
This would, effectively, replace $\bar{F}_E$ in Eq.\ (\ref{mf_tilde})
with a more complicated function of $h_E$ and $m$. Examining Eqs.\
(\ref{MF}) and (\ref{ftilde}), we see that these manipulations would
result in the following replacements,
\begin{subequations}
\begin{align}
\bar{F}_E(h_E)  & \rightarrow
\bar{F}_E(h_E, \hat{\sigma}_E(h_E, m))\nonumber\\
\bar{F}_E(h_E + \beta .m)  & \rightarrow
\bar{F}_E(h_E + \beta m, \hat{\sigma}_E(h_E, m))\nonumber
\, .
\end{align}
\end{subequations}

\noindent
Retracing
the steps that led us to Fig.\ 1d, we see that if
$\bar{F}_E(h_E, \hat{\sigma}_E(h_E, m))$ and
$\bar{F}_E(h_E + \beta m, \hat{\sigma}_E(h_E, m))$ are quasi-sigmoidal
functions of $h_E$ and $m$, we recover the nullclines in
Fig.\ 1d. Both of these
conditions are likely to hold for real neurons, since increasing
$h_E$ and $m$ correspond to increasing excitatory drive. Thus, for
neurons with reasonable gain functions, we expect Fig.\ 1d to fully
capture the shape of the nullclines.

\subsection{An example: nullclines for a simple gain function}
\label{exm-binary}

As an illustrative example, we consider a specific form for the
gain functions (the $\bar{F}_Q$), and compute the resulting nullclines
numerically. The form we choose is a rather standard one,

\begin{equation}
\bar{F}_{\k Q}(h)= \nu_{\max} \mbox{H}(h/\sigma_\k Q)
\label{tr_func}
\end{equation}

\bigskip

\noindent
where $\nu_{\max}$ is the maximum firing rate of both
excitatory and inhibitory neurons, which without loss of 
generality we take to be 100 Hz, $\mbox{H}(x)$ is given by

\begin{equation}
\mbox{H}(x)\equiv \frac{1}{1+\exp(-x)}
\, ,
\label{sigmoidalgain}
\end{equation}

\bigskip

\noindent
and $\sigma_\k Q$ is is an approximate standard deviation
based on Eq.\ (\ref{sig2}),
\begin{equation}
\sigma_{\k Q}^2 = \sum_R {K \over K_R} J^2_\k {QR} \nu^2_\k {R0}\nonumber.
\end{equation}

Before computing the nullclines for these gain functions, we
introduce a transformation that changes the nullclines without
changing the equilibria. Combining Eqs.\ (\ref{mf_tilde0_hE}) and
(\ref{mf_tilde0_m}), we see that Eq.\ (\ref{mf_tilde0_hE}) can be written
\begin{equation}
\nu_\k {E0} = \bar{F}_E(h_E) + am
\label{he_nullcline}
\, .
\end{equation}

\noindent
Note that the right hand side of Eq.\ (\ref{he_nullcline}) is an
increasing function of both $h_E$ and $m$, so the $h_E$-nullcline
based on Eq.\ (\ref{he_nullcline}) has the same qualitative
shape as the
$h_E$-nullcline based on Eq.\ (\ref{mf_tilde0_hE}). This form is more
useful than the one in Eq.\ (\ref{mf_tilde0_hE}), however, because we can
immediately write down an expression for $h_E(m)$,
\begin{equation}
h_E(m)=\bar{F}^{-1}_E(\nu_\k {E0}-am).
\label{hem}
\end{equation}

Computing the nullclines is now a straightforward numerical task, and in
Figs.\ 2a-d we plot the $m$-nullclines (green) for increasing values
of $\beta$ and the $h_E$-nullclines (blue)
for two different values of the coding level, $a$.
Because the $m$-nullcline does not depend on $a$
(see Eq.\ (\ref{mf_tilde0_m})),
there is only one $m$-nullcline in each panel.

The first thing we notice
is that when $\beta$ is sufficiently small (Fig.\ 2a), the $m$-nullcline 
consists only of a line at $m=0$, which means that
the only possible equilibria are at $m=0$,
and so retrieval states are not possible.
When $\beta$ is slightly larger 
(Fig.\ 2b), the $m$-nullcline gains a second piece away from the
line $m=0$. However, this second piece
lies below both $h_E$-nullclines, so the only intersections
are again at $m=0$, and retrieval is again not possible.
The fact that
there is no memory retrieval
when $\beta$ is small
makes sense: $\beta$ controls the
connection strength among the neurons 
within each memory, so if it is too small there will not be enough 
recurrent connectivity to produce elevated firing. 

\begin{figure}
\centering
\subfigure{\includegraphics[height=6cm,width=5.7cm]{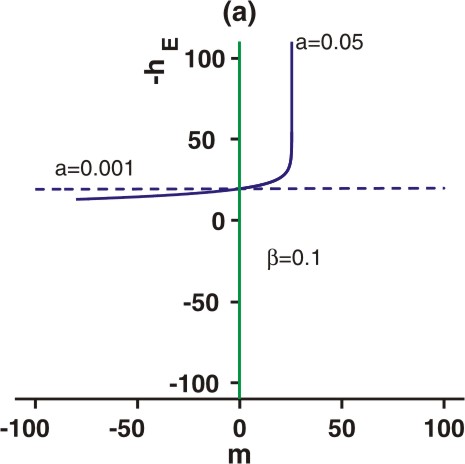}}
\subfigure{\includegraphics[height=6cm,width=5.7cm]{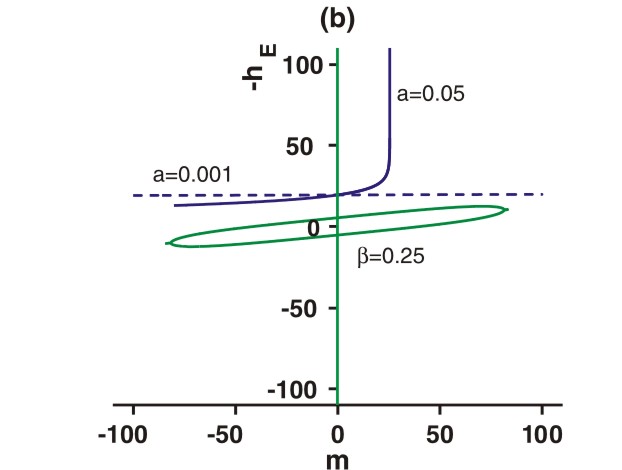}}
\subfigure{\includegraphics[height=6cm,width=5.7cm]{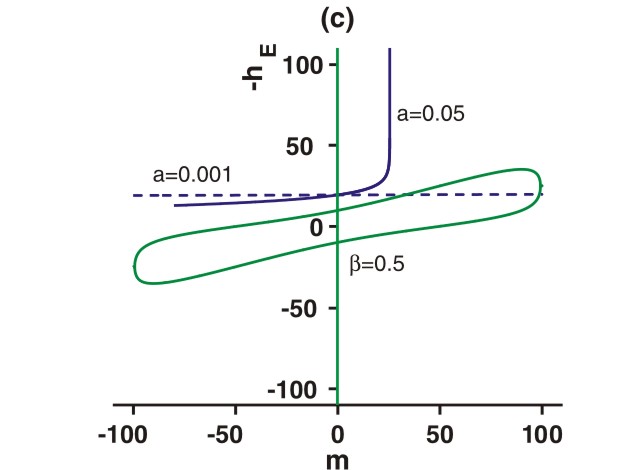}}
\subfigure{\includegraphics[height=6cm,width=5.7cm]{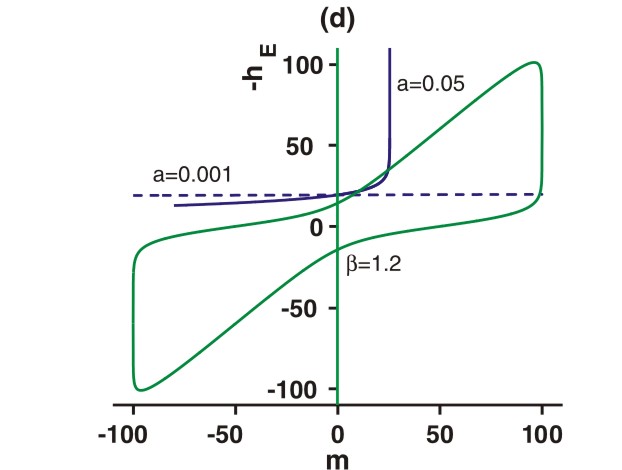}}
\caption{$h_\k E$-nullcline and $m$-nullcline for the gain function
given in Eqs.\ (\ref{tr_func}) and (\ref{sigmoidalgain}).
Different panels correspond to
different values of $\beta$, and in all
of them two $h_E$-nullclines are shown: one with $a=0.001$ (dashed blue line)
and one with $a=0.05$ (solid blue lines). The $m$-nullcline does not
depend on $a$ (Eq.\ (\ref{mf_tilde0_m})). The parameters were
$J_\k {EE}=J_\k
{IE}=1, J_\k {EI}=-1.9, J_\k {II}=-1.5, h_\k {Eex}=3, h_\k
{Iex}=2.1$, which implies, via Eqs.\ (\ref{nu0}) and (\ref{detJ}),
that $\nu_\k {E0} = 1.3$ Hz.
{\bf (a)} $\beta=0.1$. The $m$-nullcline consists only of a line at $m=0$,
so there can be no memory retrieval states.
{\bf (b)} $\beta=0.25$. The $m$-nullcline gains a second piece away
from $m=0$, but there are still no equilibria with nonzero $m$, and
thus no retrieval states.
{\bf (c)} $\beta = 0.5$ The $m$-nullcline now intersects
one of the
$h_E$-nullclines -- the one with small coding level, $a$.
{\bf (d)} $\beta=1.2$. There are now three intersections for both
values of $a$. The ones with $m=0$ and large $m$ are stable; the one
with intermediate $m$ is unstable (see Appendix III).
The $h_E$-nullcline with $a=0.001$ is essentially a straight line,
so memory retrieval occurs at a
firing rate that is too high to be biologically realistic.
The $h_E$-nullcline with $a=0.05$, on the other hand, has strong
upward curvature, so memory retrieval occurs at a much lower, and
thus biologically plausible, firing rate.
}
\label{Fig2}
\end{figure}

For still larger $\beta$, there is an intersection with one of the
$h_E$-nullclines -- the one corresponding to low coding level (Fig.\
2c). The stable equilibrium, which is the equilibrium with larger $m$,
corresponds to memory retrieval (Appendix III).
Finally, at sufficiently large $\beta$, the system acquires an
intersection with the $h_E$-nullcline corresponding to high coding
level (Fig.\ 2d). Again, the stable equilibrium is the one with larger
$m$.

An important point is that the value of $m$ at the retrieval state,
and thus the firing rate of the foreground neurons, depends strongly
on the coding level, $a$. For small $a$ (dashed blue line), retrieval
occurs near saturation, and thus at an unrealistically high firing rate.
For larger $a$ (solid blue line), the retrieval occurs at low firing
rate, consistent with experiments (when $a=0.05$ and $\beta=1.2$,
the equilibrium value of $m$ is 20 Hz). This is exactly the behaviour
we saw in the previous section.

As can be expected from these figures, increasing $\beta$ even further
would shift the intermediate intersection to negative values of $m$.
In this regime the background becomes unstable. Again this makes
sense: if the coupling among the neurons within a memory is too
strong, they become spontaneously active. 
Examining Fig.\ 1b, we see see that this occurs when the slope of
$\bar{F}_E( h_\k E + \beta m) - \bar{F}_E( h_\k E)$ with respect to
$m$ is 1 at $m=0$ (and, of course, $h_E$ is at its equilibrium value).
The value of $\beta$ at which this happens, denoted $\beta_{\max}$, is
given by
\begin{equation}
\beta_{max}
= \frac{1}{F'(F^{-1}(\nu_\k {E0}))} \nonumber,
\end{equation}

\noindent
(see Eqs.\ (\ref{mf_tilde0_m}) and (\ref{hem})).
For the sigmoidal gain function used in this example (Eq.\
(\ref{sigmoidalgain})), $\beta_{\max}$ is given by
\begin{equation}
\beta_{max}=\frac{\sigma_\k E}{\nu_\k {E0}}\left (1-\frac{\nu_\k {E0}}{\nu_{max}}\right)^{-1}
\, .
\label{max_beta}
\end{equation}

\noindent
The phase diagram for this model -- a plot showing stability and,
in the stable region, the
firing rate of the foreground neurons -- is
shown in Fig.\ \ref{bistability_region}.

\begin{figure}
\centering
\includegraphics[height=8cm,width=8cm]{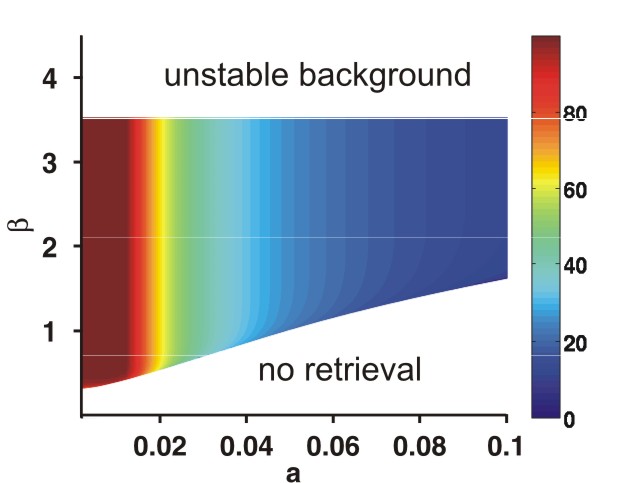}
\caption{
Phase diagram showing the values of $a$ and $\beta$ which exhibit both
a stable background and memory retrieval. The firing rate (in Hz) of the
foreground neurons is indicated by the color bar on the right.
Below the colored region only the background
exists, and it is stable. Above the colored region the
background is unstable. The upper boundary
is defined through Eq.\ (\ref{max_beta});
the lower boundary is determined numerically by finding the minimum
value of $\beta$ (for a given $a$) such that the $m$-nullcline
and $h_\k E$-nullcline intersect.
The parameters are the same as in Fig.\ 2: $J_\k {EE}=J_\k {IE}=1, J_\k
{EI}=-1.9, J_\k {II}=-1.5, h_\k {Eex}=3, h_\k {Iex}=2.1$.
}
\label{bistability_region}
\end{figure}

\subsection{Storage capacity}
\label{Str_Cap_Sec}

In the above analysis there was no way to determine how many memories
could be embedded in a network, and thus no way to determine storage
capacity. That's because we hid all effects of the quenched noise --
the noise associated with the random elements of the connectivity
matrix -- in $\bar{F}_\k E$ and $\bar{F}_\k I$ (see Eq.\ (\ref{mf_tilde})).
However, the quenched noise can have a nontrivial effect, in two ways.
First, within the context of the self-consistent signal-to-noise
analysis, it changes both $\bar{F}_\k E$ and $\bar{F}_\k I$, and thus
modifies the nullclines. Second, and potentially more important, as we
add memories we increase the number of preferred modes that can be
activated in the network, and thus we increase the quenched
noise. Either effect could cause memories to be active when
they should not be, and inactive when they should be.

To quantify these effects, we note that both scale with the
fluctuations associated with the memories that are {\em not} recalled
on any particular trial. The size of these fluctuations can be found
by computing the contribution of the memories to $\delta \hat{h}_E$,
the fluctuating piece in Eq.\ (\ref{mean_var.a}). Examining the
memory portion of the connectivity matrix, $A_{ij}$, which is given in
Eq.\ (\ref{A-ij}), and noting that $\tilde{\beta}$ is proportional to
$K^{-1}_E$ (Eq.\ (\ref{beta-scaling})), we show in Appendix II that
the variance of the quenched
fluctuations associated with this term scale as $p/K_E$ (Eq.\
(\ref{var_m})). Intuitively that is because when we sum the right hand
side of Eq.\ (\ref{A-ij}) on $j$ and $\mu$ there are $(p$$-$$1)K_E$
terms: $K_E$ that come from the sum on $j$, and $p$$-$$1$ that come
from the non-activated memories in the sum on $\mu$. Each of these
terms has variance that is $\order(1/K_E^2)$. Central limit type
arguments then tell us that the variance of such a sum is on the order
of $(p$$-$$1)K_E/K_E^2 \approx p/K_E$, where the
approximation is valid if $p$ is large.
Consequently, there is a critical value of $p/K_E$
above which none of the stored patterns could be retrieved. Thus, the
maximum number of memories in a network should scale linearly with
$K_E$. This is what we found in our simulations (Sec.\ \ref{sims}).

Unfortunately, the scale factor we found in our simulations
was small, in that the maximum number of memories scaled as $0.01 K_E$. A natural question to
ask, then, is: can the scale factor be improved by, for example, using
different parameters in our network? In the rest of this section, 
we focus on the effect of the coding level, $a$, on the storage capacity. 
We choose the coding level because, at least in simple models, 
the storage capacity is inversely proportional to $a$ \cite{Tso+88,Buh89,Tre+91}.
We have already shown that as the coding level decreases the foreground firing
rate becomes large, so we cannot make $a$ arbitrarily small. However, the minimum allowable value 
of $a$ depends on the model. What we show below, though, is that even for models which exhibit 
realistic foreground firing rate
at relatively low coding levels, the $1/a$ scaling of the storage 
capacity does not hold. This suggests that decreasing the coding level cannot be
used to increase the storage capacity in realistic networks.

Examining Eqs.\ (\ref{mean_var.a}) and (\ref{stilde}), we
see that the background neurons receive an input drawn from a Gaussian distribution
with mean $h_\k E$ and standard deviation $\hat{\sigma}_\k E$, while
the foreground neurons receive input with larger mean, $h_\k E+\beta
m$, and the same standard deviation, $\hat{\sigma}_\k E$. When the
standard deviation of these distributions, $\hat{\sigma}_\k E$, is
smaller than the separation between the means, the two populations are
well separated (Fig. \ref{Loc-Field-dist}a) and memory recall is
possible.
The standard deviation, however, is an increasing function of $p$;
see Eq.\ (\ref{MF.d}) and note that
$p$ enters this equation only through the
{\em storage load}, $\alpha$, which is defined to be
\begin{equation}
\alpha \equiv {p \over K_E}
\, .
\label{alpha}
\end{equation}

\noindent
When $\alpha$, and thus $p$, becomes large enough,
the standard deviation is of the same order 
as the separation. At this point, the two distributions
have a significant overlap with each 
other (Fig. \ref{Loc-Field-dist}b), and memory recall 
fails . 

Using this intuitive picture and Eq.\ (\ref{MF.d}), we can find the
value of $\alpha$ for which $\hat{\sigma}_\k E$ is on the order 
of the separation between the means; this should give us an
estimate of the storage capacity, $\alpha_{max}$. Using Eq.\
(\ref{MF.d}) and the fact that the means are separated by $\beta m$
(see Fig. \ref{Loc-Field-dist}), we see that this happens when 

\begin{equation} 
\left( J^2_\k {EE} + {\alpha_{max} \frac{\beta^2}{1-a}} \right)
(a \gamma_1+\gamma_2)
+
J^2_\k {EI} \gamma_3 \sim \beta^2 m^2,
\label{signa-eq-noise}
\end{equation} 

\bigskip
\noindent where

\begin{eqnarray}
&&\gamma_1=\frac{K}{K_\k E}\left[\left\langle
F_\k E^2(h_\k E + \beta m +\hat{\sigma}_\k E z) 
\right\rangle_z-\left\langle
F_\k E^2(h_\k E +\hat{\sigma}_\k E z)
\right\rangle_z\right]\nonumber\\
&&\gamma_2=\frac{K}{K_\k E}\left\langle
F_\k E^2(h_\k E +\hat{\sigma}_\k E z)
\right\rangle_z\nonumber\\
&&\gamma_3=\frac{K}{K_\k I}\left\langle
F_\k I^2(h_\k I +\sigma_\k I z) 
\right\rangle_z\nonumber.
\end{eqnarray}

\bigskip 

\noindent Solving Eq.\ (\ref{signa-eq-noise}) for $\alpha_{max}$ then leads to
 
\begin{equation}
\alpha_{max}\sim (1-a) \left(\frac{\beta^2 m^2- J^2_\k {EI} \gamma_3}{\beta^2(a \gamma_1+\gamma_2)}-\frac{J^2_\k {EE}}{\beta^2}\right). \nonumber
\end{equation}

\bigskip

If the background synaptic weights, $J_\k {EE}$ and $J_\k {EI}$, were
zero and there was zero background firing, so that $\gamma_2$
vanished, we would recover the $1/a$ scaling (in the small $a$ limit)
found in simpler models \cite{Tso+88,Buh89,Tre+91}. With nonzero
background synaptic weights, however, the capacity no longer scales as
$1/a$. Consequently, we expect that the maximum capacity cannot be improved 
much by using sparser codes.

\begin{figure}
\centering
\subfigure{\includegraphics[height=7cm,width=7.5cm]{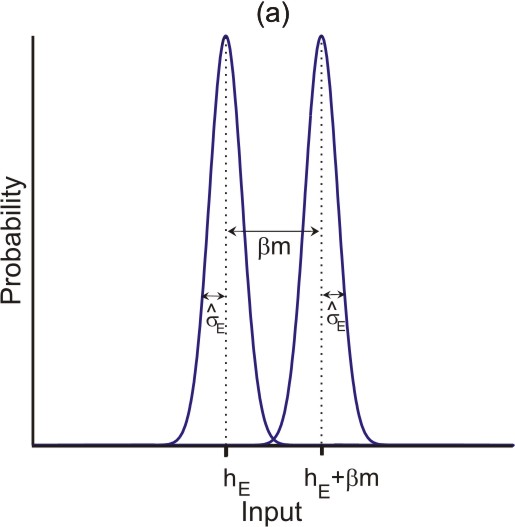}}
\subfigure{\includegraphics[height=7cm,width=9cm]{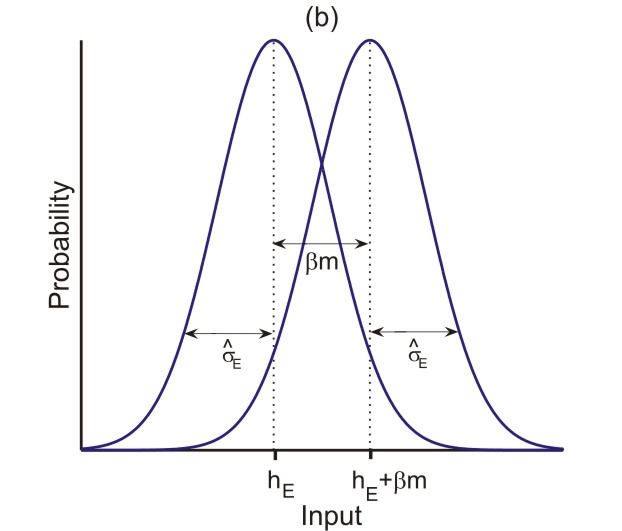}}
\caption{Distribution of inputs to foreground (mean=$h_E$)
and background (mean=$h_E + \beta m$) neurons,
and its relation to storage capacity. Both inputs
have a Gaussian distribution. The means are separated by
$\beta m$ and the standard deviation of both distributions is
$\hat{\sigma}_\k E$. {\bf (a)} The standard deviation is much smaller than the
distance between the means of the two distributions. In this regime, the
two populations are well separated, there is no interference between 
them, and memory retrieval is supported.
{\bf (b)} As $\alpha$ increases, $\hat{\sigma}_\k E$ also increases 
(Eq.\ (\ref{MF.d}))
while $m$ changes rather slowly (Eq.\ (\ref{MF.b})), so
the distributions start to overlap. When 
the overlap becomes large, noise dominates the signal, and memory 
recall is no longer possible. 
The value of $\alpha$ at which this happens is the storage capacity,
$\alpha_{max}$.
}
\label{Loc-Field-dist}
\end{figure}

\section{Computer simulations}
\label{sims}

Our mean field analysis gave us two predictions. The first is
that if the background synaptic weights, the $\tilde{J}$, scale as
$K^{-1/2}$, the foreground weights, $A$, scale as $K^{-1}$, and the
coding level, $a$, is sufficiently high, then both the background and
foreground neurons should operate in the balanced regime and the
neurons should fire irregularly. The second is that the number of
memories that can be stored is proportional to
the number of excitatory connections per neuron, $K_\k E$. 

To test these predictions, we perform simulations
with large networks of spiking neurons. We start by finding, for a
particular network size, parameters such that both foreground
and background neurons exhibit irregular activity. We then increase
the size of the network while scaling the synaptic weights according
to the above prescriptions. If the larger networks continue to
exhibit irregular activity, then our predicted scalings are
correct. To test the relation between storage capacity and
number of connections per neuron, we calculate the storage capacity
for networks with different sizes. A linear relation would
indicate a scaling consistent with our predictions.

\subsection{Network model}

Each neuron is modeled as a conductance-based quadratic
integrate and fire (QIF) neuron. Dendritic trees
and axonal arborizations are not considered. The spikes generated in
any neuron immediately affect all the postsynaptic neurons connected
to it. The membrane potential of neuron $i$ of type $Q$,
denoted $V_{Qi}$ evolves according to 

\begin{subequations}
\begin{align}
\tau {dV_{Qi} \over dt} & =
{(V_{Qi} - V_r)(V_{Qi} - V_t) \over V_t - V_r} + V_{0i}
\\
\nonumber
&- (V_{Qi} - {\cal E}_E) \sum_{j \in E} \tilde{J}^{QE}_{ij} s_{ij} (t)
- (V_{Qi} - {\cal E}_I) \sum_{j \in I}
\tilde{J}^{QI}_{ij} s_{ij} (t)+
\tilde{h}_{Q ex,i}
\\
{d s_{ij} \over dt} & = -{s_{ij} \over \tau_s} +
\sum_k \delta(t - t_j^k)
\, .
\end{align}
\label{dvdt}
\end{subequations}

\noindent
Here $\tau$ is the membrane time constant, $\tau_s$ is the synaptic
time constant, $V_r$ and $V_t$ are the nominal
resting and threshold voltages, $V_{0i}$ determines
the actual resting and threshold voltages,
($V_{0i}$ is constant for each $i$, but as a function of $i$ it's a
Gaussian random variable with mean $V_0$ and standard deviation
$\Delta V_0$),
$\tilde{J}^{QR}_{ij}$ is the connection strength from cell $j$ in
population $R$ to cell $i$ in population $Q$, ${\cal E}_E$ and
${\cal E}_I$ are the excitatory and inhibitory reversal potentials,
respectively, the notation $j \in R$ means sum over only those cells
of type $R$, $\delta(\cdot)$ is the Dirac $\delta$-function, $t_j^k$
is the $k^{\rm th}$ spike emitted by neuron $j$, and $\tilde{h}_{Qex,i}$
is the external input to neuron $i$ of type $Q$. The external input
is modeled as
\begin{subequations}
\begin{align}
\tilde{h}_{Qex,i}&=(V_{Qi} - {\cal E}_E) \tilde{J}_\k {Qex} s_\k {Qex}(t) \label{ext_input_a}\\
{d s_\k {Qex}\over dt}&= -{s_\k {Qex} \over \tau_s} +
\sum_k \delta(t - t^k_\k {Qex}),
\end{align}
\label{ext_input}
\end{subequations}

\noindent
where the ${t^k_\k {Qex}}$ are the times of the external spikes. 
These are taken to be Poisson at constant rate $\nu_\k {Qex}$.

There are two features of these equations that are worth commenting
on. First, the connection strengths, $\tilde{J}^{QR}_{ij}$, are
completely analogous to the ones given in Eq.\ (\ref{loc-field}).
Thus, although the $\tilde{J}^{QR}_{ij}$ in Eq.\ (\ref{dvdt})
have different numerical values than those in Eq.\ (\ref{loc-field}),
they should have the same scaling with connectivity \cite{Shriki03,Lerch04,Hertz04}. The same is also
true of $\tilde{h}_{Qex}$, except that here
$\tilde{h}_{Qex}$ has temporal fluctuations whereas in Eq.\ (\ref{loc-field})
it does not.
Second, we have included a term $V_{0i}$,
which has the effect of making the
resting membrane potential and threshold of each cell different. This
was not explicitly modeled in our mean field analysis, although
it would not have made much difference -- it would have only added to
the quenched noise.

The $\tilde{J}_{ij}^{QR}$ have the same form as in Eq.\ (\ref{syn-weights}), 
except that we introduce an extra scaling factor so that connection 
strengths can be directly related to PSP size.  Specifically, we use the fact
that if neuron $j$ spikes and neuron $i$
is at rest, then the PSP generated at neuron $i$ will have
peak amplitude $\tilde{J}^{QR}_{ij} V_R$ where
\begin{equation}
V_R =
{{\cal E}_R - V_r
\over
(\tau / \tau_s)
\exp [ \ln ( \tau / \tau_s ) / (\tau / \tau_s - 1)]}\nonumber
\,; 
\end{equation}

\bigskip

\noindent
see \cite{Lat00} for a derivation of this expression. This suggests
that we should scale our connection strengths by $V_R$, so we write

\begin{equation}
\tilde{J}^{QR}_{ij}={c^{QR}_{ij} \over V_R} \,
\left[
\tilde{J}_{QR} +
\delta_{Q,E} \delta_{R,E} \,
\tilde{\beta} \sum_{\mu=1}^p \xi_i^\mu (\xi_j^\mu - a)
\right]
\label{J}
\end{equation}

\bigskip

\noindent
where $c^{QR}_{ij}$ is the same binary random variable defined in Eq.\
(\ref{prob_c}), $\delta_{Q,R}$ is the Kronecker delta, and $\tilde{J}_{QR}$ and $\tilde{\beta}$ in Eq.\
(\ref{J}) correspond to, but typically have different numerical values than, the
ones in Eqs.\ (\ref{syn-weights}) and (\ref{beta-scaling}).
If $V_R$ is in mV, then $\tilde{J}_{QR}$ is the
peak postsynaptic potential, in mV, that occurs in a neuron in pool $Q$ when a neuron in pool
$R$ fires (assuming the two  are connected, the postsynaptic neuron is
at rest, and $\tilde{\beta}=0$).

Our analytical results have been derived by assuming current based
neurons. However, it is possible to extend such analysis to a more
realistic network of conductance based neurons by noting that the
effective connection strength in a conductance based model is
proportional to the PSP size \cite{Shriki03,Lerch04,Hertz04}. Thus,
for the network to operate in the balanced regime, we should have the
following scalings,

\begin{subequations}
\begin{align}
\tilde{J}_{QR} & \sim K^{-1/2},
\\
\tilde{J}_{Qex} & \sim K^{-1/2}, 
\label{J-ex-scaling}
\\
p & \sim K,
\\
\tilde{\beta} & \sim K^{-1}
\, .
\end{align}
\label{scaling}
\end{subequations}

\noindent
Note that the mean external excitatory input must be proportional to $K^{1/2}$. Therefore,
given Eq.\ (\ref{ext_input_a}) and the scaling of $\tilde{J}_{Qex}$ in Eq.\ (\ref{J-ex-scaling}),
the firing rate of the neurons that provide external input, $\nu_{Qex}$, should scale as $K$.

We performed simulations using three different networks, called
Networks 1, 2 and 3, that differ in the number of neurons (they
contain a total of 10,000, 20,000 and 30,000, respectively).
In all three networks
$c=0.15$, so $K$ is proportional to the total number of neurons in the
network. Because of the scaling in Eq.\ (\ref{scaling}),
the values of $\tilde{J}_{QR}$,
$\tilde{\beta}$, $\nu_{Qex}$ and $p$ also differ.
The parameters for the three networks are given in Table I.
Our goal in these simulations is to determine whether, as predicted by
our mean field analysis, the above scaling leads to behaviour that is
independent of $K$ and the firing of both foreground and background
neurons is irregular. 

\begin{table}
{{\bf Table I.}  Parameters used in the simulations. $\tilde{\beta}_{predicted}$}
(the only parameter not actually used in the simulations) is the predicted value of 
$\tilde{\beta}$ based on our mean field analysis.
\medskip

\begin{tabular}{|l|c|c|c|} \hline
\bf{Network number} & 1 &2 &  3\\ \hline
Excitatory neurons & 8,000 & 16,000 & 24,000 \\ \hline
Inhibitory neurons & 2,000 & 4,000  & 6,000 \\ \hline
$K (\equiv K_E + K_I)$ & 1,500 & 3,000  & 4,500 \\ \hline
$\tilde{J}_{EE}$  & 0.5 mV & 0.35 mV &0.29 mV\\ \hline
$\tilde{J}_{EI}$  & 1.0 mV &  0.71 mV&0.58 mV\\ \hline
$\tilde{J}_{IE}, \tilde{J}_{II}$ & -4.0 mV& -2.83 mV& -2.31 mV \\ \hline
$\tilde{J}_{Eex}$  & 0.5 mV & 0.35 mV &0.29 mV\\ \hline
$\tilde{J}_{Iex}$  & 1.0 mV &  0.71 mV&0.58 mV\\ \hline
$\nu_\k {Eex}$ & 1000 & 2000 & 3000\\ \hline
$\nu_\k {Iex}$ & 450 & 900 & 1350\\ \hline
$\tilde{\beta}$ & 0.168  & 0.101 & 0.077\\ \hline
 $\tilde{\beta}_{\rm predicted}$
   &  -- & 0.083 & 0.056\\ \hline
$p$ & 5 & 10 & 15\\ \hline
$V_{0}$ & 1.5 mV & 1.5 mV & 1.5 mV\\ \hline
$\Delta V_{0}$ & 0.5 mV & 0.5 mV & 0.5 mV\\ \hline
$a$ & 0.1 & 0.1& 0.1\\ \hline
$c$ & 0.15& 0.15& 0.15\\ \hline
$\tau$ & 10 ms& 10 ms& 10 ms\\ \hline
$\tau_s$ & 3 ms& 3 ms& 3 ms\\ \hline
$V_r$ & -65 mV & -65 mV& -65 mV\\ \hline
$V_t$ & -50 mV & -50 mV&-50 mV\\ \hline
${\cal E}_E$ & 0 mV & 0 mV& 0 mV \\ \hline
${\cal E}_I$ & -80 mV &-80 mV& -80 mV\\ \hline
Time step & 0.5 ms & 0.5 ms& 0.5 ms\\ \hline
\end{tabular}
\end{table}

\subsection{Building a balanced network}

Our first step in assessing our mean field predictions is to
build a network that operates in the balanced regime and supports
retrieval states. To test whether a network is
operating in the balanced regime, we rely on two indicators. One is
that it exhibits
irregular firing, quantified by the coefficient of variation (CV) --
the ratio of the standard deviation to the mean interspike
interval -- and that the CV is independent of K. The second is that the mean excitatory and inhibitory firing
rates scale linearly with the external input, as predicted by
Eq.\ (\ref{nu0}). To test whether a network
supports retrieval states, we simply activate a memory by bombarding
all the neurons within a memory with excitatory input,
and ask whether the memory stays active for several seconds.
Very little fine tuning was required to find a network that
exhibited both balance and retrieval states: we simply chose
reasonable peak PSPs, set the coding level, $a$, to 0.1, and increased
$\tilde{\beta}$ until at least one memory was stored.

In Fig.\ \ref{Fig6}a we show an example of the retrieval of a stored
pattern for Network 1. The first 2 seconds in this figure consists
of background firing; at $t=2$ seconds, neurons selective for one of
the patterns receive an excitatory external input lasting for $100$
ms; and at $t=27.3$ seconds, the same neurons receive an inhibitory
external input, which again lasts for 100 ms. The blue line is
the mean firing rate of the foreground neurons, the 
black line is the mean firing rate of the excitatory neurons
(both foreground and background) and the red line is the
mean firing rates of the inhibitory neurons.

Two points are worth mentioning. One is that the background
firing rate in our simulations is lower than the background firing
rate observed in studies of delay activity, which range from
$1.5$ to $8$ Hz \cite{Nakamura95}, although we
should point out that the firing rates determined from extracellular
recordings may be overestimated due to selection bias \cite{Lennie03}. We could, however,
achieve a higher background rate by increasing the excitatory
external input; an
example is shown in Fig.\ \ref{high_back},
for which the network parameters are the same
as Network 1 (Fig.\ 5a) except that the external input to excitatory and
inhibitory neurons is five times higher, $\beta$ is a factor
of about two higher,
and there is just one
stored pattern instead of five. With the higher input,
the background and foreground rates are in the range reported from
neurons in, for example, anterior
ventral temporal cortex \cite{Miy+88,Miy88} and entorhinal cortex
\cite{Nakamura95}.

The second point is that during retrieval
the mean firing rates of the excitatory and inhibitory
neurons differ from the background rates; i.e., from the
rates when no memories are
activated. This appears to be
inconsistent with the balance condition,
which predicts that the mean firing rate during the activation
of a memory is the same as that when the network is in the background
state (see Eq.\ (\ref{nu0})).
However, this prediction holds only in the limit of infinite
connectivity. For finite connectivity, there are corrections,
and they are
particularly important when the firing rate is low \cite{Van98}.
For example,
in Fig.\ \ref{Fig6}a the average excitatory activity
increased from 0.28 Hz in the background to 1.07 Hz
during retrieval
(an increase of about 400\%), whereas
in Fig.\ \ref{high_back}, where the background is higher,
it increased from 1.06 Hz
to 1.73 Hz (an increase of 60\%).
Thus, the increase in the mean
excitatory firing rate during retrieval is reduced when the
firing rate is higher. However, this is accompanied, at least in
the parameter range we looked at, by a decrease in the storage
capacity. Since we
would like to study the scaling of storage capacity, we operate
in the lower firing rate regime.
A detailed search of
parameter space is required to determine whether
both high storage capacity and high background firing
can be achieved.

\begin{figure}
\centering
\subfigure{\includegraphics[height=6cm,width=6cm]{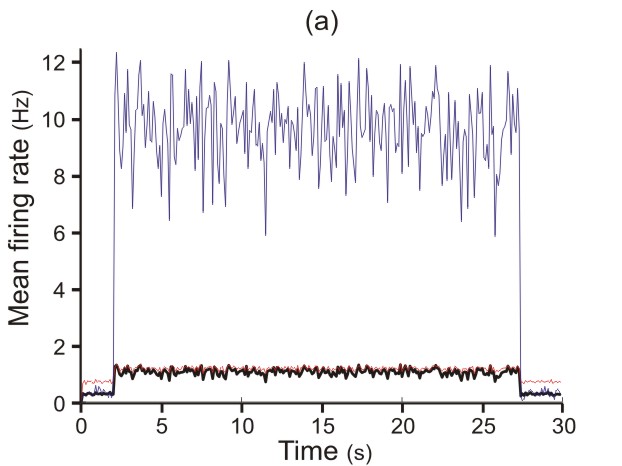}}\\
\subfigure{\includegraphics[height=6cm,width=6cm]{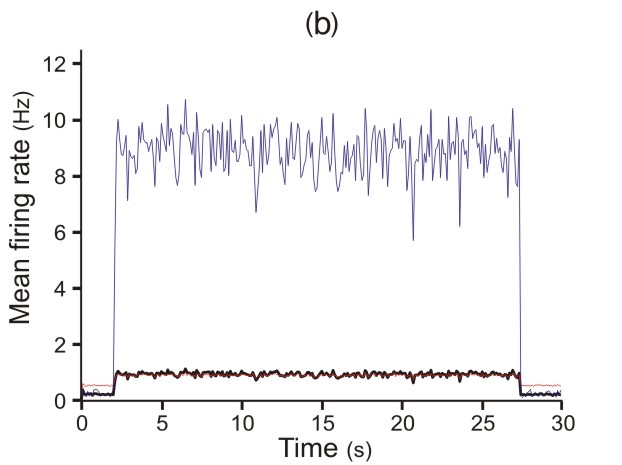}}\\
\subfigure{\includegraphics[height=6cm,width=6cm]{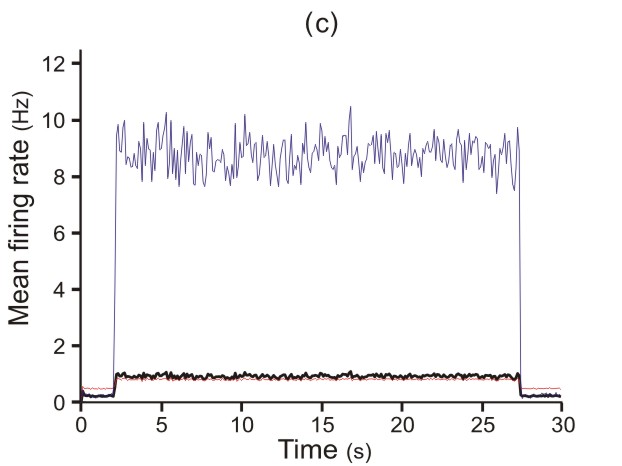}}
\caption{Examples of activation of a retrieval state. {\bf (a)} Network 1. 
{\bf (b)} Network 2. {\bf (c)} Network 3. Colors indicate mean
population activity. Blue: foreground
neurons. Black: excitatory neurons. Red: inhibitory neurons.
At $t=2$ seconds, neurons selective for one of the patterns receive a
$100$ ms barrage of excitatory input;
at $t=27.3$ seconds, the same neurons
receive a barrage of inhibitory input.}
\label{Fig6}
\end{figure}

\begin{figure}
\centering
\includegraphics[height=6cm,width=6cm]{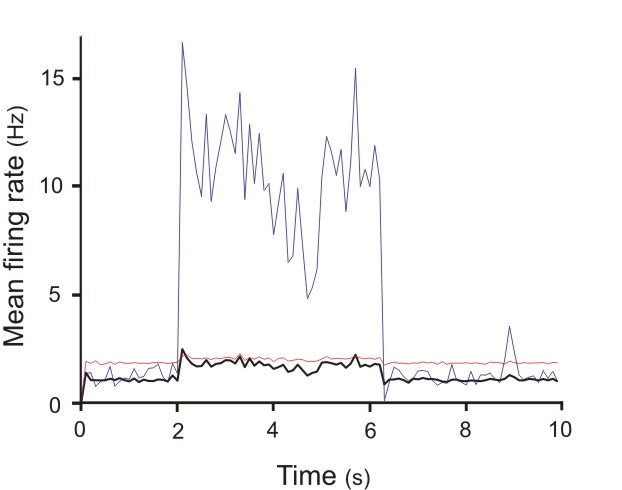}
\caption{Retrieval states with higher external input than in
Fig.\ \ref{Fig6}, and thus higher background firing rate.
All parameters except
$\nu_\k {Eex}, \nu_\k {Iex}, \beta$ and $p$
are the same as in Network 1: here
$\nu_\k {Eex}=5000$ Hz, $\nu_\k {Iex}=2250$ Hz, $\beta=0.325$
and $p=1$, versus Network I, where
$\nu_\k {Eex}=1000$ Hz, $\nu_\k {Iex}=450$ Hz, $\beta=0.167$
and $p=5$
The stored pattern receives input for
$100$ ms, starting at
$t=2$ seconds, and then receives an external inhibitory current,
again for $100$ ms, starting at $t=6.2$ seconds. 
}
\label{high_back}
\end{figure}

In Fig.\ \ref{Fig7}a we show the CV versus firing rate, again for
Network 1. Here and in what follows, the CV is calculated only for
those neurons that emit at least 5 spikes during the 25 second
period that the pattern is active. The data in this figure fall into
two clusters, one (blue dots) corresponds to background neurons and the
other (red crosses) to foreground neurons. The distributions of CVs and
firing rates are shown in Figs.\ \ref{Fig7}b and c.
The CV of both background and
foreground neurons are on the order of $0.8$, which indicates
irregular firing. This suggests that the network is operating
in the balanced regime. To further test for balance,
in Fig.\ \ref{Fig8}a we plot the average excitatory
and inhibitory firing rates versus the external input. As predicted by
Eqs.\ (\ref{balanced_cond}) and (\ref{nu0}), the relation is
approximately linear.

\begin{figure}
\centering
\subfigure{\includegraphics[height=5cm,width=5cm]{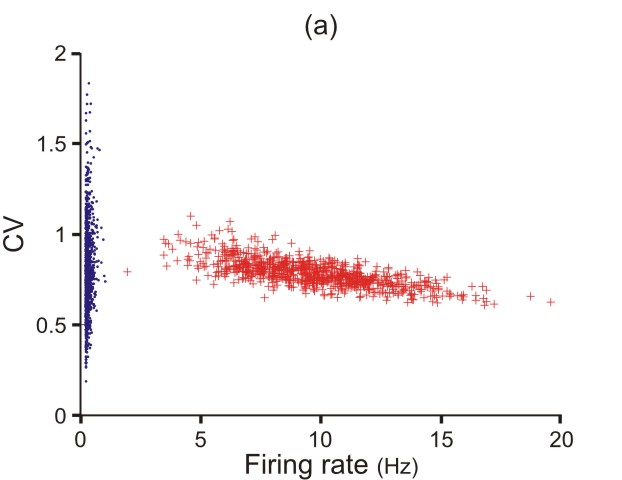}}
\subfigure{\includegraphics[height=5cm,width=5cm]{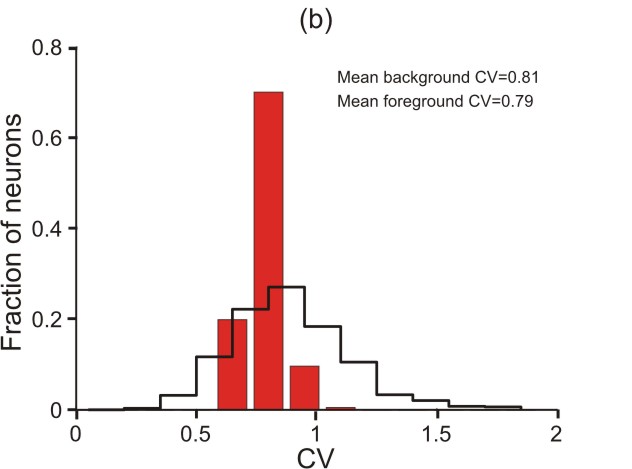}}
\subfigure{\includegraphics[height=5cm,width=5cm]{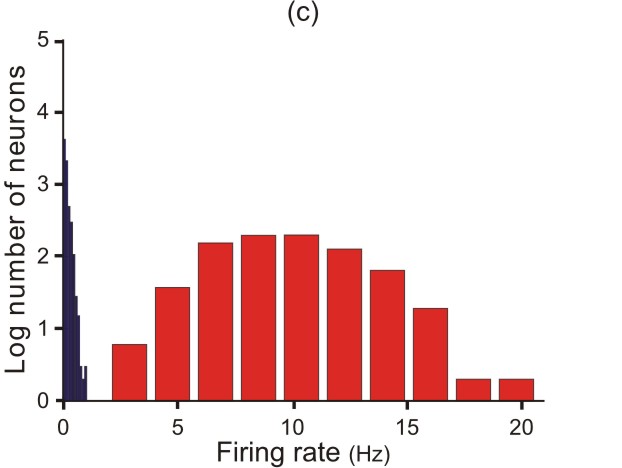}}\\
\subfigure{\includegraphics[height=5cm,width=5cm]{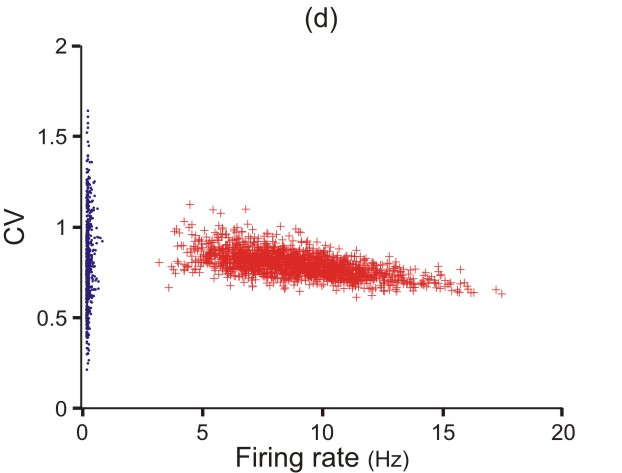}}
\subfigure{\includegraphics[height=5cm,width=5cm]{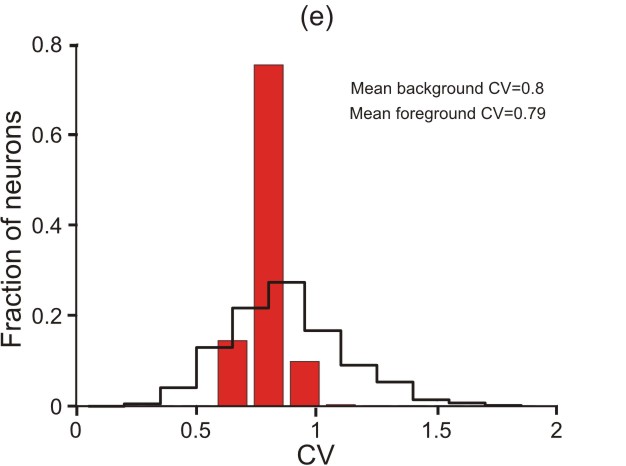}}
\subfigure{\includegraphics[height=5cm,width=5cm]{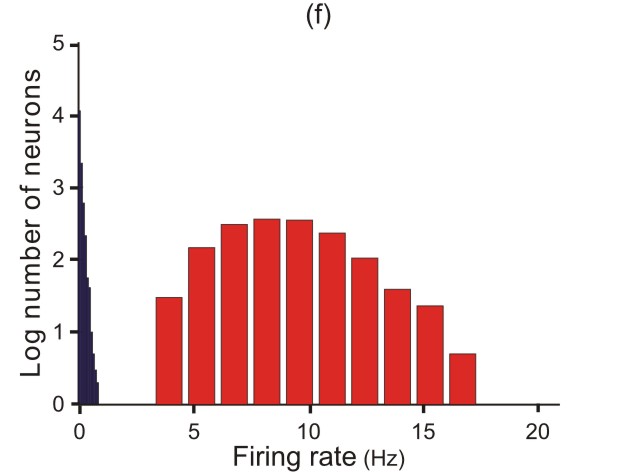}}\\
\subfigure{\includegraphics[height=5cm,width=5cm]{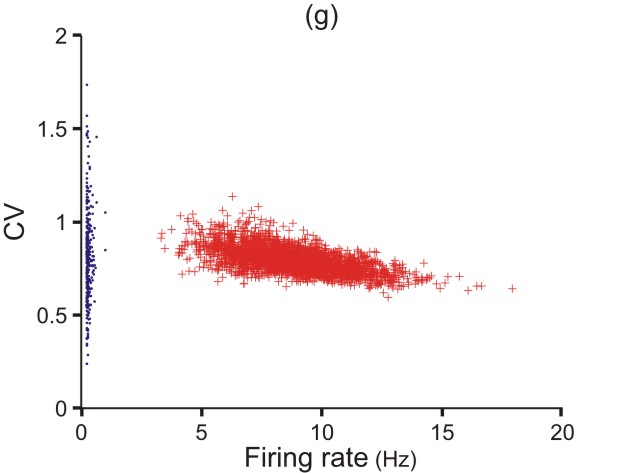}}
\subfigure{\includegraphics[height=5cm,width=5cm]{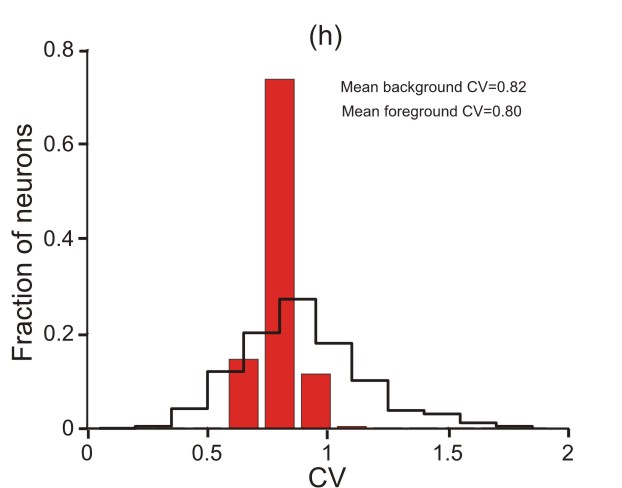}}
\subfigure{\includegraphics[height=5cm,width=5cm]{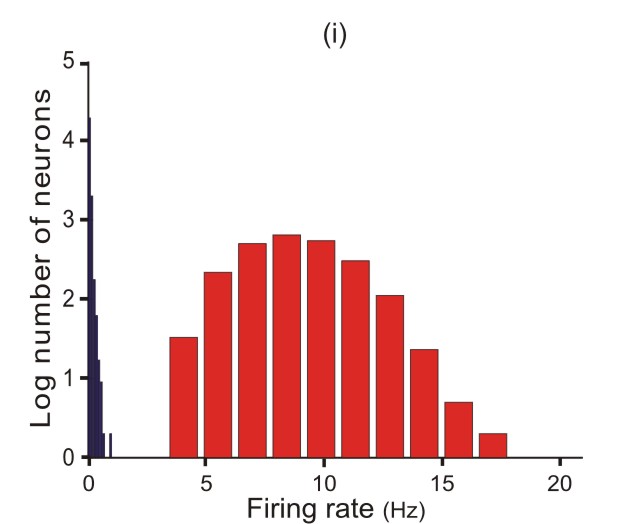}}\\
\caption{The distribution of CVs (coefficients of variation) 
and firing rates for foreground and 
background neurons. The first, second and third rows correspond to 
Networks 1, 2 and 3, respectively.
{\bf Column 1.} CV versus the firing rate of 
background (blue dots) and foreground (red crosses) 
neurons. Consistent with activation of a memory state, the neurons 
fall into two clusters, one corresponding to the foreground
and the other to the background. {\bf Column 2.} Distribution 
of CVs for foreground (filled red bars) 
and background (solid line).
The mean of both distributions is about $0.8$, reflecting the fact that 
the neurons are firing irregularly. 
{\bf Column 3.} Distribution of firing 
rates for foreground (filled red bars) and background (solid line).}
\label{Fig7}
\end{figure}

\begin{figure}
\centering
\includegraphics[height=9cm,width=9cm]{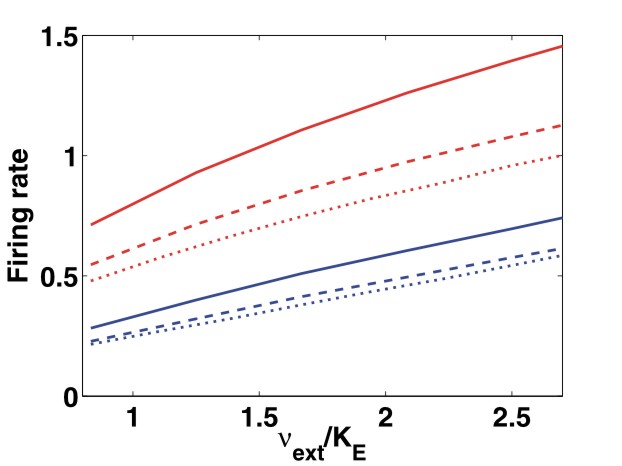}\\
\caption{Average excitatory (blue) and inhibitory (red) firing rate
versus external input to excitatory neurons, measured as firing rate per connection $(\nu_{\k E ex}/K_\k {E})$.
The ratio $\nu_{\k I ex} / \nu_{\k E ex}$ was fixed at 0.45. Full lines, dashed lines
and dotted lines correspond to Networks 1, 2 and 3, respectively. The
average rates are calculated during a 4 second period which consists
of background firing only. The linear relationship between the mean
inhibitory and excitatory firing rates and the external input is a
signature of the balanced regime. 
}
\label{Fig8}
\end{figure}
\subsection{Scaling of the parameters}

To test our predicted scaling with the number of connections, we
considered networks with two and three times the number of
neurons and connections as in Network 1; these are Networks 2 and 3.
At the same time we scaled 
$\tilde{J}_{QR}$ by $K^{-1/2}$, $\nu_\k {Qex}$ by $K$, and $p$ by $K$
(see Eqs.\ (\ref{scaling}a-c)).
The value of $\tilde{\beta}$ was set, as in Network 1, to
the minimum value that results in retrieval
of a single stored pattern. The $1/K$ scaling of $\beta$ (Eq.\ (\ref{beta-scaling})) gives us the 
values reported as $\tilde{\beta}_{predicted}$ in Table I. The values
found from simulations ($\tilde{\beta}$ in Table I) do not
exactly follow the expected $1/K$ scaling: $\tilde{\beta}$ is 20\% too
large in Network 2 and 40\% too large in Network 3. 
As discussed in Appendix IV, this is because of finite $K$ effects,
and the trends we see here follow the trends predicted in that
appendix.

Examples of stored memories in Networks 2 and 3 are shown in Figs.\
\ref{Fig6}b and c, the CV versus firing rate is shown in Figs.\
\ref{Fig7}d and g, and the distribution of background and foreground
CV and firing rates during the 25 second period that the memory is
active are shown in Fig.\ \ref{Fig7}e and f for Network 2 and Fig.\
\ref{Fig7}h and i for Network 3. These plots show that when the
connection strengths are scaled properly, both the background and
foreground neurons exhibit irregular firing, just as in Network 1.
Finally, Fig.\ \ref{Fig8}b and c show the relationship between the
external input and the firing rate of the inhibitory and excitatory
populations. As we saw for Network 1, the firing rate of excitatory
and inhibitory neurons are linearly related to external input,
further evidence for the balanced regime. In theory the lines should 
lie on top of each other, however due to finite size effects this does not
happen. The fact that finite size effects are responsible for this deviation
from the theory can be seen by noting that the lines corresponding to Network 2 and 
Network 3 are much closer to each other than Network 1 and Network 2.

\subsection{Scaling of the maximum number of memories}

Our last prediction is that the maximum number of memories should be
linear in the number of excitatory connections, $K_E$. To test this,
for each of our three networks we increased the number of patterns,
$p$, until the network failed to exhibit retrieval states.
Specifically, we performed simulations as describe in Fig.\
\ref{Fig6}, except that the memory was active for 6 seconds rather
than 25. For each value of $p$, we activated all $p$ memories one at a
time. If the mean activity of the foreground neurons during the 6
seconds of activation was at least 3 times larger than the
activity averaged over all excitatory neurons, then that memory was
said to be successfully retrieved.
 
The results of these simulations are shown in Fig.\ \ref{Fig9}a, where
we plot the fraction of successful retrievals versus $p/K_{E}$ for
the three networks. Consistent with our predictions, the transition to
a regime where none of the patterns could be retrieved occurs at
approximately the same value of $p/K_{E}$ for all three networks.
Moreover, as one would expect, the transition for the largest network
is sharper than for the others. 

Although Fig.\ \ref{Fig9}a shows that $p_{\max}$ scales linearly
with $K_E$, in these simulations $N_E$ also scales with $K_E$,
so this does not rule out the possibility that $p_{\max}$ is
proportional to $N_E$ rather than $K_E$.
To test for this, in Fig.\ \ref{Fig9}b we plot
the fraction of successful retrievals versus $p/K_E$, but this time
with $K_E$ fixed and $N_E$ varied. This figure shows that
$p_{\max}$ is proportional to $K_E$, not $N_E$, ruling out the $N_E$
scaling.

\begin{figure}
\centering
\subfigure{\includegraphics[height=7.2cm,width=8cm]{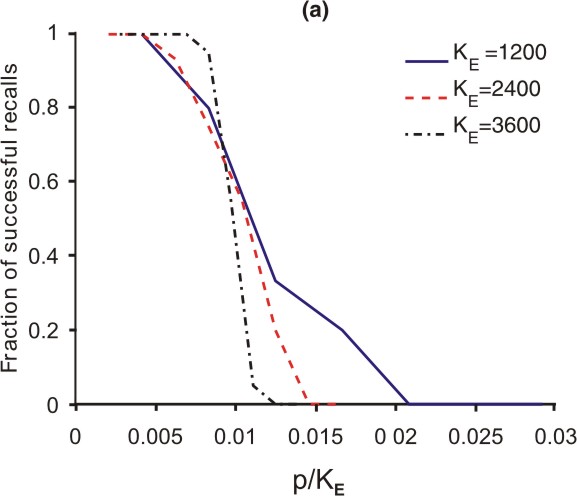}}
\subfigure{\includegraphics[height=7.2cm,width=8cm]{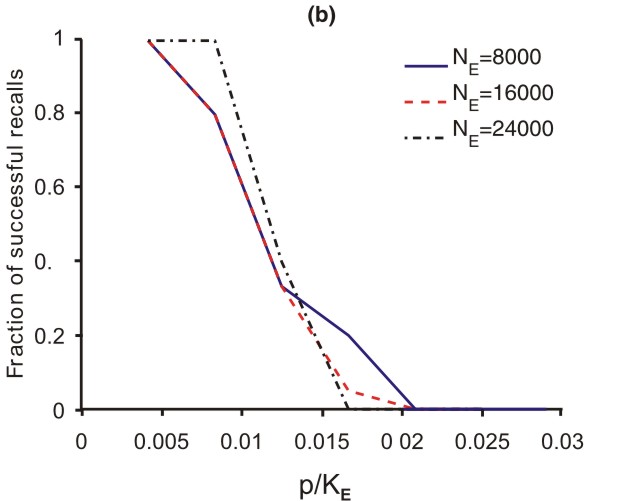}}
\caption{Scaling of the maximum number of patterns with the number 
of excitatory connections per neuron, $K_\k E$. {\bf (a)} The fraction of 
successful runs versus  
the storage load, $\alpha=p/K_\k E$, for three different values of $K_\k E$.
The size of the network is scaled such that we always have $K_\k E/N_\k E=K_\k I/N_\k I=0.15$. 
There is a critical value 
of $\alpha$, above which the fraction of successful 
runs is zero; this is the storage capacity $\alpha_{max}$. 
The transition at $\alpha_{max}$
is sharp for $K_\k E=3600$ but smoother for $K_\k E=2400$ and
$K_\k E=1200$, due to finite size effects. The fact that $\alpha_{max}$
is almost the same for all three values of $K_\k E$ implies that 
the maximum number of patterns that could be stored and retrieved,
$p_{\max}$, is linear in $K_\k E$.
{\bf (b)} The fraction of successful runs versus
the storage load, $\alpha=p/K_\k E$, for three networks with all
parameters, except for the total number of neurons in the network, 
equal to those of Network 1.
This figure shows that increasing the size of the network does not
change $p_{\max}$.
}
\label{Fig9}
\end{figure}

\section{Discussion}

In this paper we addressed two questions. First, can all the
neurons in an attractor network -- both background and foreground --
exhibit irregular firing? And second, what is the storage capacity in
networks of realistic spiking neurons? To answer these
questions, we applied self-consistent signal-to-noise analysis to
large networks of excitatory and inhibitory neurons, and we
performed simulations with spiking neurons to test the predictions of
that analysis.

Our primary finding is that two conditions must be met to guarantee
irregular firing of both foreground and background neurons.
The first is proper scaling with the number of
connections per neuron, $K$: the strength of the
background weight matrix must
scale as $K^{-1/2}$ and the strength of the
structured part of the weight matrix (the part responsible for
the memories) as $K^{-1}$.
What this scaling does is guarantee ``balance,'' meaning the
network dynamically adjusts its firing rates so that the mean input to a
neuron is on the same order as the fluctuations, independent of $K$.
This in turn guarantees that the degree of irregular firing is
independent of $K$.

While balance is a necessary condition for irregular firing,
it is not sufficient. That's because
balance ensures only that the mean and fluctuations are independent of $K$,
but does not rule out the possibility that the mean is much
larger than the fluctuations, which would result in regular firing.
To ensure that this does not happen, a second
condition must be satisfied: the coding level,
$a$, must be above some ($K$-independent) threshold. This condition is needed to
ensure that the coupling between background and foreground neurons
is sufficiently strong to stabilize a low firing rate foreground state
on the unstable branch of the $m$-nullcline (see Fig.\ 1).

The analysis that led to predictions of irregular firing also
quite naturally provided us with information about the capacity of
attractor networks -- the maximum number of patterns that could be
stored and successfully retrieved. What we found, under very general conditions,
was that this maximum, denoted $p_{\max}$, is linear in the number of
excitatory connections per neuron, $K_E$. This scaling relation has
been observed in studies of simplified attractor
networks \cite{Somp86,Der+87,Tre+91}, but, as discussed in
the Introduction, those models did not include
all the features that are necessary for a realistic recurrent networks.
Thus, the analysis performed here is the first to show that the
number of memories is linear in $K_E$ in
biophysically plausible networks.

\subsection{Scaling in other models, and the importance of $\order(1)$
input to the foreground neurons}
Note that there are other types of scaling, different from what
we proposed, which can result in irregular firing of both
foreground and background neurons. What is critical is that the net
input a foreground neuron receives from the other foreground neurons
should be $\order(1)$. We achieved this by letting the structured part
of the connection matrix
(the second term in Eq.\ (\ref{loc-field-separated.a})) be
$\order(1/K)$ and using a coding level, $a$, that was $\order(1)$.
%
However, this is not the only possible combination
of connection strengths and
coding levels, and in the two other
studies that address both scaling and irregularity in memory networks
\cite{Van05,Renart06}, different combinations were used.
In the model proposed by van
Vreesjwik and Sompolinsky \cite{Van05}, the structured
part of their connection matrix was a factor of $K^{1/2}$ larger
than ours; to balance that,
the coding level was a factor of $K^{1/2}$ smaller.
In the model proposed by Renart et al.\ \cite{Renart06}, the
structured part of the synaptic weights was $K$ times larger than
ours, so their
coding level had to scale as $\order(1/K)$. Whether such low coding
levels are consistent with reality needs further investigation;
however, data from studies conducted on selectivity of neurons to
visual stimuli suggests that it is too low \cite{Rol95}. In addition
to the very low coding level that these two models require, they also
exhibit non-biologically high foreground firing rate.
Nevertheless, the model of Renart et al.\ \cite{Renart06} does have
one advantage over others: the foreground neurons are as
irregular as, or even more irregular than, the background neurons,
something our model does not achieve (see
next section).

\subsection{Not as irregular as it could be}
\label{not_as_irr}
Although our simulations showed irregular activity, we found
that the mean CV was only about $0.8$. This is smaller than the values
measured {\em in vivo}, which are normally close to, or slightly
above, one \cite{Noda70,Burn76,Soft93,Hol96,Comp03}. In addition, in
our simulations the CV showed a small, but consistent, decrease with
firing rate (see the first column in Fig.\ 7). This is due to the fact
that with the scaling that we chose, the fluctuations in the input
current to foreground and background neurons is the same but the mean
currents to the foreground neurons is higher (Appendix I). This
decrease in the CV disagrees slightly with a study by Compte et al.\
\cite{Comp03}, who found that the CV in prefrontal cortex does not
depend on the mean firing rate, at least in a spatial memory task.
While there are many possible reasons for this discrepancy, a likely
one arises from the fact that the neurons in our network contained
only two time scales, the membrane and synaptic time constants, and
both were short: 10 ms for the former and 3 ms for the latter. Real
neurons, however, have a host of long time scales that could
contribute to irregularity \cite{Dingledine99}. In addition, {\em in
vivo} optical imaging \cite{Tsodyks99,Kenet03,Yuzhi06} and
multi-electrode \cite{Fiser04} studies indicate that the background
activity varies coherently and over long time scales, on the order of
seconds, something we did not model. Both of these would increase the
CV, although how much remains to be seen.

Although multiple time scales could certainly increase irregularity,
it is not the only possible way to do this. As discussed in the Introduction and in the previous section, 
the model proposed by Renart et al \cite{Renart06} also increases irregularity, and is 
consistent with the experimental results of
Compte et al \cite{Comp03}. However, it requires a very small coding
level ($a \sim 1/K$), and fine tuning of the parameters.

\subsection{Subthreshold versus suprathreshold persistent activity}

In conventional models of persistent activity \cite{Ami97,Bru00,BrunelWang01},
the foreground activity necessarily lies on the concave part of the excitatory 
gain function, $F_\k E(h_\k E)$, whereas the background activity lies on the convex part. 
Since the inflection point of realistic gain functions is typically near the 
firing threshold \cite{Fourcaud02,Bru03}, this type of bistability is called suprathreshold
bistability \cite{Bru00,Renart06}. Because the concave part of the gain function is
typically at high firing rate, with suprathreshold bistability it is 
hard to have either low foreground firing rate or high CV. Consequently, there has been 
interest in understanding whether it is possible to have 
subthreshold bistability; that is, whether it is possible for both foreground and background solutions to lie on 
the subthreshold part of the gain function \cite{Renart06}.}

The model presented here can in fact show subthreshold bistability: as discussed in 
Sec. 2.3, increasing the coding level, $a$, brings the foreground firing rate very close to the
background rate. Therefore, for sufficiently large $a$, the foreground state would be on
the convex part of the transfer function. Our model, and the recently proposed
model by Renart et al \cite{Renart06}, are the only ones
that can show subthreshold bistability.

\subsection{Bimodal distribution of firing rates}

One rather striking feature of our networks is that they all produce a
highly bimodal distribution of firing rates: as can be seen in the
first and third columns of Fig.\ 7, the background neurons fire at a
much lower rate than the foreground neurons --
so much lower, in fact, that they
form a distinct, and easily recognizable, population. This occurs
because the patterns we store -- the $\xi_i^\mu$ --
are {\em binary}, which makes the average
input current to every neuron in the foreground exactly the same.
This feature is potentially problematic, as the distinction between foreground
and background rates observed in experiments is not nearly as striking
as the one in Fig.\ 7 \cite{Fuster82}. However, this feature is not essential
to our analysis, for two reasons. First, as discussed in Sec.\
3.2 (see especially Fig.\ 6b), we deliberately made the background
firing rate low to increase the capacity. Second, it is also easy to extend our analysis to real valued patterns in
which the elements of the $\xi_i^\mu$ are drawn from a continuous
distribution \cite{Tre+91}. Under this, more realistic, scenario, it it should
be possible to match the statistics of the response seen in the
cortex. This will be the subject of future work.

\subsection{Fine tuning of the weights}

In our model, every time a new patterns is learned, the weights change by
an amount proportional to $K^{-1}$. This is a factor of $K^{-1/2}$
smaller than the background weights. Since weight
changes are unlikely to be under such fine control, it is natural to
ask whether errors during learning will lead to a major reduction in
storage capacity. The answer, of course, depends on the size of the errors. 
In Appendix V, we show that errors can be larger than the weight changes by a factor of
$(K/p)^{1/2}$, with only a small change in storage capacity.
More specifically, every time a pattern is learned, noise of
$\order\left((Kp)^{-1/2}\right)$ can be added to the synaptic strength, and the
network will retain its ability to store and recall patterns.

Although this result tells us that the noise in the weight changes can
be large compared to the structured part, the fine tuning problem is
not entirely eliminated: the noise must still be a factor of
$p^{1/2}$ smaller than the background weights. Because of the low
storage capacity found in these networks (at most 2.5\% \cite{Lat04}), 
even when $K$ is as large as $10,000$, $1/p^{1/2}$ is
on the order of 6\%. It seems plausible that biological machinery has evolved to
achieve this kind of precision. However, for
networks with larger capacity, the requirements on the precision of
the weight would be more stringent.

It is also possible to have a probabilistic learning rule for which
the changes in the weights are on the same order as the background weight, but this decreases 
the capacity significantly by a factor of $\sqrt{K}$ (see Appendix V, Eq.\ (\ref{prob-learning}); 
we thank an anonymous reviewer for pointing this out).
Although this probabilistic learning rule guarantees a balanced state
with irregular background and foreground firing, it has the drawback that the storage capacity
scales as $\sqrt{K}$ rather than $K$.

\subsection{Low storage capacity}

Although we showed that $p_{\max}\propto K_E$, we did not compute
analytically the constant of proportionality. In our simulations, this
constant was small: from Fig.\ \ref{Fig9}, $p_{\max}$ is about
$0.01K_E$, which means that for $K_E=10,000$ we can store only about
100 patterns. It is important, though, to note that we made no attempt to optimize our 
network with respect to other parameters, so
the constant of proportionality 0.01 is unlikely to be a fundamental
limit. In fact, Latham and Nirenberg \cite{Lat04} were able to store
about 50 patterns in a network with $2000$ excitatory connections, 2.5
times larger than our capacity. Interestingly, the only substantial
difference between their network and ours was that in theirs the
background activity was generated by endogenously active neurons
rather than external input.

Can we further increase the scaling factor? One
potential mechanism is to decrease the coding 
level, $a$, since, at least in
simple models \cite{Tso+88,Buh89,Tre+91},
the maximum number of patterns that could be stored and retrieved 
is inversely proportional to the coding level. But, as we
showed in Sec.\ \ref{Str_Cap_Sec}, realistic networks do {\em not}
exhibit this $1/a$ scaling. Consequently, sparse coding cannot be used
as a way to improve the storage capacity in our network. Simplified models also 
suggest that one can increase the storage capacity by a factor of 3--4
by using other schemes, such as non-binary patterns \cite{Tre+91}, or spatially 
correlated patterns \cite{Monasson93}. Whether these techniques can be extended to the kind of 
network we have studied here is not clear, and requires further investigation. However, 
an increase beyond a factor of 3--4, to a capacity above around $0.1$, seems unlikely
within this class of networks.

In any case, there is a limit to the number of memories that can be stored in a
single attractor network with a fixed number of connections per
neuron, no matter how many neurons in the network. This suggests that,
in order to make the best use of the existing connections,
realistic working memory systems must be composed of interconnected
modules. In this paradigm, each module would consist of an attractor
network \cite{Okane92,Lev+99,Kro+05}. Such 
modular structure naively suggests a combinatorial increase in
storage capacity; however, understanding how to 
achieve such an increase has proved difficult. For simple 
models whose storage capacity could be calculated 
analytically, either no increase in the storage capacity 
\cite{Okane92} or a modest increase \cite{Kro+05} was found. 
It is yet to be determined how modular networks could 
be implemented in realistic networks of spiking neurons, 
and what their storage capacity would be.

\renewcommand{\theequation}{I-\arabic{equation}}
\setcounter{equation}{0}
\section*{Appendix I: Fast fluctuations}

The starting point for essentially all of our analysis is
Eq.\ (\ref{dynamic_fr}), which, when combined with Eq.\
(\ref{loc-field}), tells us that the
time evolution of the firing rate of each neuron is a function
purely
of the firing rates of the other neurons.
At a microscopic
level, though, each neuron sees as input a set of spikes,
not rates. However, for our model, rate-based equations do apply,
as we show now.

In a spiking, current-based network, the input, $h_{Qi}(t)$, to the
$i^{th}$ neuron in population $Q$ has the form
\begin{subequations}
\begin{align}
h_{Qi}(t) &=\sum_R \sum_{j\in R} \tilde{J}^{QR}_{ij} S_{j}(t)
\label{local-field-time.a}
\\
S^{\k R}_{j} &=\sum_k f_\k R(t-t^k_{j}),
\label{local-field-time.b}
\end{align}
\end{subequations}

\noindent
where $t^{k}_{j}$ is the time of the $k^{th}$ spike on the $j^{th}$
neuron,
$f_\k R(t)$, which mimics the PSP, is a non-negative function that
integrates to 1 and vanishes for $t < 0$ and $t$ large (greater
than a few 10s of ms).
In a slight departure from our usual convention, $R$ can refer
to external input ($R=ex$) as well as excitatory and
inhibitory input ($R=E,I$).

Our first step is to divide the input, $h_{Qi}$,
into a mean and a temporally fluctuating piece.
The mean, which is found by time averaging the right hand side of Eq.\
(\ref{local-field-time.a}) and using the fact that $f_\k R(t)$
integrates to 1, is simply
\begin{equation}
\langle h_\k {Qi}(t)\rangle_t=\sum_R \sum_{j\in R} \tilde{J}^{QR}_{ij} \langle S^{\k R}_{j}(t)\rangle_t=\sum_R \sum_{j\in R} \tilde{J}^{QR}_{ij} \nu_{Rj},
\label{hbar_t}
\end{equation}

\noindent
where $\langle \cdots \rangle_t$ represents a temporal average.
The temporally fluctuating piece of the input can then be written
\begin{subequations}
\begin{align}
&\delta h_{\k {Qi}}(t)=h_{\k {Qi}}(t)-\langle
h_{\k {Qi}}(t)\rangle_t=\sum_R \sum_{j\in R} \tilde{J}^{QR}_{ij} \delta
S^\k R_{j}(t),
\label{deltah}
\\
&\delta S^\k R_{j}(t)=S^\k R_{j}(t)-\nu_{Rj}.
\end{align}
\end{subequations}

\noindent
The fluctuations, $\delta h_{{\k Q}i}$, have zero mean by
construction, and their correlation function, $C_{Qi}(\tau)$, is
defined to be  
\begin{equation}
C_{Qi}(\tau)=\langle \delta h_{Qi}(t+\tau)\delta h_{Qi}(t)\rangle_t.
\end{equation}

Assuming that $h_{Qi}$ is Gaussian (which is reasonable if there are
a large number of neurons and they are not too correlated), then the
firing rate depends only in the mean, $\langle h_{Qi}(t)\rangle_t$,
and the correlation function, $C_{Qi}(\tau)$.
If the correlation function is independent of $i$, then the
only $i$-dependence in the
firing rate is through the mean input, and we recover Eq.\
(\ref{dynamic_fr}). What we now show is that, for our
model, $C_{Qi}$ does not depend on $i$.

To understand the behaviour of $C_{Qi}$, we
express it in terms of
$\delta S^\k R_j(t)$; using Eq.\ (\ref{deltah}), we have
\begin{equation}
C_{Qi}(\tau) =
\sum_{R,R'}\sum_{j\in R,j'\in R'} \tilde{J}^{QR}_{ij} \tilde{J}^{QR'}_{ij'}
\langle \delta S^\k R_j(t) \delta S^\k R_{j'}(t+\tau) \rangle_t
\nonumber
\end{equation}

\noindent
Under the assumption that the neurons are very weakly correlated, only
the terms with $j=j'$ survive, and this expression simplifies to
\begin{equation}
C_{Qi}(\tau) =
\sum_{R} \sum_{j\in R} (\tilde{J}^{QR}_{ij})^2
\langle \delta S^\k R_j(t) \delta S^\k R_{j}(t+\tau) \rangle_t
\, .
\nonumber
\end{equation}

\noindent
Let us focus on the sum on $j$ on the right hand side of this
expression. For $Q \ne E$ or $R \ne E$, this sum is given by
(see Eqs.\ (\ref{syn-weights}b-d))
\begin{equation}
\sum_{j\in R} (\tilde{J}^{QR}_{ij})^2
\langle \delta S^\k R_j(t) \delta S^\k R_{j}(t+\tau) \rangle_t
=
\tilde{J}_{QR}^2 \sum_{j\in R} c_{ij}^{QR}
\langle \delta S^\k R_j(t) \delta S^\k R_{j}(t+\tau) \rangle_t
\, .
\label{corr_qr}
\end{equation}

\noindent
For sparsely connected networks, $c_{ij}^{QR}$ is independent of
$\delta S^\k R_j(t)$.
Consequently, we can replace $c_{ij}^{QR}$ on the right hand side of
Eq.\ (\ref{corr_qr}) by its average, $c$,
and the right hand side becomes independent of $i$.

For $Q=R=E$ the situation is more complicated, as
$\tilde{J}_{ij}^{EE}$ has an additional dependence on $A_{ij}$, the
structured part of the connectivity. Specifically using
Eq.\ (\ref{syn-weights}a)
and again replacing $c_{ij}^{EE}$ by its average, $c$,
we have
\begin{equation}
\sum_{j\in E} (\tilde{J}^{EE}_{ij})^2
\langle \delta S^\k E_j(t) \delta S^\k E_{j}(t+\tau) \rangle_t
=
c \sum_{j\in E}
(\tilde{J}_{EE} + A_{ij})^2
\langle \delta S^\k E_j(t) \delta S^\k E_{j}(t+\tau) \rangle_t
\, .
\label{corr_ee}
\end{equation}

\noindent
As discussed in Appendix II, $A_{ij}$ receives contributions from two
sources: the
$p-1$ patterns that are not activated, and the one pattern that is. The
non-activated patterns are not correlated with $\delta S_j$, so they
can be averaged separately in Eq.\ (\ref{corr_ee}), and thus do not produce
any $i$-dependence. The activated pattern, on the other hand is
correlated with $\delta S_j$. However, the connection strength for the
one activated pattern is smaller than $\tilde{J}_{EE}$ by a factor of
$K^{-1/2}$ (see Sec.\ 1).
Consequently, in the high connectivity limit, we can
ignore this contribution, and the right hand side of Eq.\ (\ref{corr_ee})
is independent of $i$. This in turn implies that $C_{Qi}$
depends only on $Q$.

The upshot of this analysis is that the only $i$-dependence in the firing
rate comes from $\langle h_{Qi}(t)\rangle_t$. Moreover, comparing
Eqs.\ (\ref{loc-field}) and (\ref{hbar_t}), we see that
$\langle h_{Qi}(t)\rangle_t$ is exactly equal to $h_{Qi}$,
the input current to the firing rate function, $F_Q$, that appears in
Eq.\ (\ref{dynamic_fr}). Thus, for the model used here, the rate-based
formulation is indeed correct. What we do not do is compute
$F_Q$, as that would require that we compute the correlation
function, $C_Q(\tau)$, self-consistently, which is nontrivial
\cite{Hertz04}. However, our results depend
very weakly on the precise form of $F_Q$, so it is not necessary to
have an explicit expression for it.

\renewcommand{\theequation}{II-\arabic{equation}}
\setcounter{equation}{0}
\section*{Appendix II: Mean-field equations}

In this appendix we derive the mean field equations for the model
described in Sec.\ \ref{Model}. As discussed in the main text, the
derivation of these equations revolves around finding the distributions
of $\delta \hat{h}_{Ei}$ and $\delta h_{Ii}$, 
the fluctuations around the mean
excitatory and inhibitory synaptic input (both quantities are defined implicitly
in Eqs.\ (\ref{loc-field-separated}-\ref{he_hi})).
The main assumption we make is that $\delta\hat{h}_\k{Ei}$
and $\delta h_\k{Ii}$ are zero mean Gaussian random variables, so 
all we need to do is find their
variances self-consistently. In addition, primarily for simplicity
(and because it is reasonable in large networks in the brain), we
assume that the number of connections is small compared to the number
of neurons, so $c \ll 1$.

Our first step is to simplify the expressions for our main order
parameters, $\nu_E$, $m$, and $\nu_I$. In the context of the
self-consistent signal-to-noise analysis, ``simplify'' means ``replace
sums by Gaussian integrals''. To see how to do this, note that, for
any function $g$,
\begin{equation}
{1 \over N_E} \sum_{i=1}^{N_E} g(\delta \hat{h}_{Ei}) \approx
\int Dz \, g \left({\rm Var}[\hat{h}_E]^{1/2} z \right)\nonumber
\end{equation}

\noindent
where Var$[\cdot]$ indicates variance, exact
equality holds in the $N_E \rightarrow \infty$ limit
(but approximate equality typically holds when $N_E$ is only a few
hundred), and
\begin{equation}
Dz \equiv {dz \, e^{-z^2/2} \over (2 \pi)^{1/2}}\nonumber
\, .
\end{equation}

\noindent
A similar expression applies, of course, to $\delta h_{Ii}$.

Applying the sum-goes-to-integral rule to Eq.\ (\ref{mf_sum}), we have

\begin{subequations}
\begin{align}
\nu_E &= \int Dz \,
\left\langle F_E \left( h_E + \xi \beta m +
{\rm Var}[\delta \hat{h}_E]^{1/2} z \right)\right\rangle_\xi
\\
m &= \int Dz \,
\left\langle
{\xi - a \over a(1-a)}
F_E\left( h_E + \xi \beta m + {\rm Var}[\delta \hat{h}_E]^{1/2} z \right)
\right\rangle_\xi
\\
\nu_I &= \int Dz \,
F_I\left( h_I + {\rm Var}[\delta h_I]^{1/2} z \right)
\end{align}
\label{mf_int}
\end{subequations}

\noindent
where the average over $\xi$ is with respect to the probability
distribution given in Eq.\ (\ref{prob_a}).

To complete Eq.\ (\ref{mf_int}), we need the variance of
$\delta \hat{h}_{Ei}$ and $\delta h_{Ii}$.
It is convenient to break
the former into two pieces,
$\delta \hat{h}_{Ei} = \delta h_{Ei} + \delta h_{mi}$, where
the first, $\delta h_{Ei}$, is
associated with the background neurons, and the second,
$\delta h_{mi}$, is
associated with the foreground neurons
(both will be defined shortly).
Then, examining Eqs.\ (\ref{loc-field-separated}-\ref{overlap}),
and performing a small amount of algebra,
we find that

\begin{subequations}
\begin{align}
\delta h_{Ei} & =
{K^{1/2} \over K_E} J_{EE} \sum_j (c^{\kl EE}_{ij} - c) \nu_{Ej} +
{K^{1/2} \over K_I} J_{EI} \sum_j (c^{\kl EI}_{ij} - c) \nu_{\k Ij}
\\
\delta h_{Ii} & =
{K^{1/2} \over K_E} J_{IE} \sum_j (c^{\kl IE}_{ij} - c) \nu_{\k Ej} +
{K^{1/2} \over K_I} J_{II} \sum_j (c^{\kl II}_{ij} - c) \nu_{\k Ij}
\end{align}
\label{var_back}
\end{subequations}

\noindent
and
\begin{equation}
\delta h_{mi} =
{\beta \over K_E  a(1-a)} \sum_j
\sum_\mu
(c^{\kl EE}_{ij} - c \delta_{\mu,1})
\xi_i^\mu (\xi_j^\mu - a) \nu_{\k Ej}
\, .
\label{var_fore}
\end{equation}

\noindent
Here $\delta_{\mu, \nu}$ is the Kronecker delta; it is 1 if
$\mu=\nu$ and zero otherwise. In addition, for notational convenience,
we have returned the superscript ``1'' to $\xi_i$. For the rest of the
appendix, we will use 
$\xi_i^1$ and $\xi_i$ interchangeably.

Let us focus first on the contribution from the background,
Eq.\ (\ref{var_back}). Since
$c_{ij}^{\kl QR}$ is equal to $c$ on average, the mean of
both terms on the right hand side of
Eq.\ (\ref{var_back}) is zero. Moreover, these terms are uncorrelated,
so their variances add. The variance of the $QR^{\rm th}$ term is then

\begin{equation}
{\rm Var} \left[
{K^{1/2} \over K_R} \sum_j (c^{\kl QR}_{ij} - c)
\nu_{\k Rj}
\right]
=
{K \over K_R^2} \sum_{jj'}
\left\langle (c^{\kl QR}_{ij} - c)
(c^{\kl QR}_{ij'} - c) \right\rangle
\nu_{\k Rj} \nu_{\k Rj'} \nonumber
\end{equation}

\noindent
where the angle brackets represent an average over the distribution of
$c_{ij}^{\kl QR}$.
Because $c_{ij}^{\kl QR}$ and $c_{ij'}^{\kl QR}$ are independent when
$j \ne j'$, only terms with $j \ne j'$ produce a nonzero average.
Thus, all we need is the variance of $c_{ij}^{QR} - c$, which is
given by

\begin{equation}
{\rm Var} \left[ c^{\kl QR}_{ij} - c \right]
=
c (1 - c) \approx c \nonumber
\end{equation}

\noindent
(the last approximation is valid because, as mentioned above, we are
assuming $c \ll 1$).
Performing the sums over $j$ and $j'$ and collecting terms, we have

\begin{equation}
{\rm Var} \left[
{K^{1/2} \over K_R} \sum_j (c^{\kl QR}_{ij} - c)
\nu_{\k Rj}
\right]
=
{K \over K_R}
{1 \over N_R} \sum_j \nu_{\k Rj}^2
\equiv
{K \over K_R}
\langle \nu_{\k R}^2 \rangle
\, .
\label{varQ3}
\end{equation}

\noindent
The term on the right hand side,
$\langle \nu_{\k R}^2 \rangle$, is the
second moment of the firing rate of the neurons in pool $R$.
Inserting Eq.\ (\ref{varQ3}) into (\ref{var_back}), we find
that
\begin{equation}
{\rm Var} [ \delta h_Q ] \equiv
\sigma_Q^2 =
\sum_R
{K \over K_R}
J_{QR}^2 \langle \nu_\k R^2 \rangle
\, .
\label{sig2}
\end{equation}

The last quantity we need is the variance of $\delta h_m$. A naive
approach to computing it proceeds along lines similar to those
described above: assume all the terms in the sum over $j$ and $\mu$
in Eq.\ (\ref{var_fore}) are
independent, so that the variance of $\delta h_m$ is just $pN_E$ (the
number of terms in the sum) times the variance of each term. This
yields, with rather loose notation for averages and
ignoring the $\order(K_E^{-1})$ correction associated with $\mu=1$,
\begin{equation}
{\rm Var}[\delta h_m] = {\beta^2 p N_E\over K_E^2 a^2(1-a)^2}
\, 
\langle (c^{\kl EE}_{ij})^2 \rangle
\langle \xi^2 \rangle
\langle (\xi-a)^2 \rangle \langle \nu_{\k E}^2 \rangle
\, .\nonumber
\end{equation}

\noindent
All the averages in this expression are straightforward:
$\langle (c^{\kl EE}_{ij})^2 \rangle = c$,
$\langle \xi^2 \rangle = a$,
$\langle (\xi-a)^2 \rangle = a(1-a)$, and
$\langle \nu_{\k E}^2 \rangle$ was defined in Eq.\ (\ref{varQ3}).
Putting all this together and defining $\rho^2$ to be the variance of
$\delta h_m$, we have
\begin{equation}
{\rm Var}[\delta h_m] \equiv \rho^2 =
{\beta^2 \, \langle \nu_{\k E}^2 \rangle \over 1-a}
\, {p \over K_E}
\, 
\, .
\label{var_m}
\end{equation}

While Eq.\ (\ref{var_m}) turns out to be correct,
our derivation left out a potentially important
effect: correlations between the patterns, $\xi^{\mu}_i$, and the
overlaps, $m^\mu_i$ (the latter is defined in Eq.\ (\ref{ss1}) below).
These correlations, which
arise from the recurrent feedback,
turn out to scale as $c$, and so can be neglected
\cite{Der+87,Evans89,Roudi04}. Rather than
show this here, we delay it until the end of the
appendix (see subsection ``{\em Loop corrections vanish in the small
$c$ limit}'').

To write our mean field equations in a compact form, it is convenient
to define the total excitatory variance,
\begin{equation}
\hat{\sigma}_E^2 \equiv \sigma_E^2 + \rho^2
\, .
\label{stilde}
\end{equation}

\noindent
Then, combining Eqs.\ (\ref{steady_state}),
(\ref{mf_int}), (\ref{sig2}) and (\ref{var_m}),
the mean field equations become

\begin{subequations}
\begin{align}
\nu_\k E &= \Big\langle
\langle F_E(h_E + \xi \beta m + \hat{\sigma}_E z) \rangle_{\xi}
\Big\rangle_z
\label{MF.a}
\\
m &= \Big\langle
\langle (a(1-a))^{-1} (\xi-a)
F_E(h_E + \xi \beta m + \hat{\sigma}_E z) \rangle_{\xi}
\Big\rangle_z
\label{MF.b}
\\
\nu_\k I &= \left\langle F_I(h_I + \sigma_I z) \right\rangle_z
\label{MF.c}
\\
\hat{\sigma}_E^2 &=
{K \over K_E} \left( J^2_{EE} + {\alpha \beta^2 \over 1-a} \right)
\left\langle
\langle F_E^2(h_E + \xi \beta m +\hat{\sigma}_E z) \rangle_{\xi}
\right\rangle_z
+
{K \over K_I} J^2_{EI} \left\langle F_I^2(h_I+\sigma_I z) \right\rangle_z
\label{MF.d}
\\
\sigma_I^2 &=
{K \over K_E} J^2_{IE}
\left\langle
\langle F_E^2(h_E + \xi \beta m +\hat{\sigma}_E z) \rangle_{\xi}
\right\rangle_z
+
{K \over K_I} J^2_{II} \left\langle F_I^2(h_I+\sigma_I z) \right\rangle_z
\label{MF.e}
\end{align}
\label{MF}
\end{subequations}

\noindent
where the subscript $z$ indicates a Gaussian average,
\begin{equation}
\langle (\cdot) \rangle_z \equiv \int Dz \, (\cdot)
\, ,\nonumber
\end{equation}

\noindent
and, recall, $\alpha = p / k_E$ (Eq.\ (\ref{alpha})).

Finally, it is convenient to explicitly perform the averages over
$\xi$ that appear in Eq.\ (\ref{MF}). Defining
\begin{equation}
\bar{F}^{(k)}_R (h, \sigma) \equiv \int Dz \, F^k_R(h, \sigma z),
\label{ftilde}
\end{equation}

\noindent
the relevant averages become

\begin{subequations}
\begin{align}
&
\left\langle
\langle F_E^k(h_E + \xi \beta m + \hat{\sigma}_E z) \rangle_{\xi}
\right\rangle_z
=
(1-a) \bar{F}^{(k)}_E (h_E, \hat{\sigma}_E) +
a \bar{F}^{(k)}_E (h_E + \beta m, \hat{\sigma}_E)
\label{MFx.a}
\\
&
\left\langle
\langle (a(1-a))^{-1} (\xi-a)
F_E(h_E + \xi \beta m + \hat{\sigma}_E z) \rangle_{\xi}
\right\rangle_z
=
\bar{F}_E (h_E + \beta m, \hat{\sigma}_E) -
\bar{F}_E (h_E, \hat{\sigma}_E)
\label{MFx.b}
\, .
\end{align}
\label{MFx}
\end{subequations}

\noindent
The functions $\bar{F}_E$ and $\bar{F}_I$ that we used in
Eq.\ (\ref{mf_tilde}) are equivalent to the ones defined in
Eq.\ (\ref{ftilde}), although in the main text
we suppressed the dependence on the
standard deviation and dropped the superscript.

Equation (\ref{MF}) constitutes our full set of mean field equations.
A key component of these equations is that the number of memories,
$p$, enters only through the variable $\alpha$, which is $p/K_E$. Thus,
the number of memories that can be embedded in a network of this type
is linear in the number of connections.

\bigskip
\noindent
{\em Loop corrections vanish in the small $c$ limit}

\noindent
To correctly treat the loop corrections in our derivation of the
variance of $\delta h_m$, we need to be explicit about the
correlations induced by the patterns, $\xi_i^\mu$. We start by
defining the $i$-dependent overlap, $m^\mu_i$, as
\begin{equation}
m^{\mu}_i = {1 \over K_E a(1-a)}
\sum_j (c^{\kl EE}_{ij} - \delta_{\mu, 1} c)
(\xi^{\mu}_j-a) \nu_{\k Ej}
\, .
\label{ss1}
\end{equation}

\noindent
Inserting this into Eq.\ (\ref{var_fore}) leads to
\begin{equation}
\delta h_{mi} =
\beta \sum_\mu \xi_i^\mu m^\mu_i
\, .
\label{var1}
\end{equation}

\noindent
Each of the terms $m^\mu_i$ is a Gaussian random variable whose
variance must be determined self-consistently. This can be done by
inserting Eq.\ (\ref{steady_state}) into Eq.\ (\ref{ss1}) to
derive a set of nonlinear equations for the $m^\mu_i$.
There are two types of terms to consider: the activated 
memory, for which $\mu=1$, and the non-activated 
memories, for which $\mu\neq1$. However, in  the large $p$ 
limit we can safely ignore the one term corresponding to $\mu=1$.
Thus, considering the contributions from memories with $\mu \ne 1$, 
we have
\begin{equation}
m^{\mu}_i = {1 \over K_E a(1-a)}
\sum_j c^{\kl EE}_{ij} (\xi^{\mu}_j-a)
F_E \left(h_E + \xi^1_j \beta m^1 + \delta h_{Ej} +
\beta \xi_j^\mu m^\mu_j +
\beta \sum_{\nu \ne \mu,1} \xi_j^\nu m^\nu_j \right)\nonumber
\, .
\end{equation}

\noindent
Taylor expanding around $m^\mu_j = 0$ and defining

\begin{subequations}
\begin{align}
F^\mu_{E j} & \equiv
F_E \left(h_E + \xi^1_j \beta m^1 + \delta h_{Ej} +
\beta \sum_{\nu \ne \mu} \xi_j^\nu m^\nu_j \right)\nonumber
\\
F^{\mu'}_{Ej} & \equiv
F_E' \left(h_E + \xi^1_j \beta m^1 + \delta h_{Ej} +
\beta \sum_{\nu \ne \mu,1} \xi_j^\nu m^\nu_j \right)\nonumber
\end{align}
\end{subequations}

\noindent
where a prime denotes a derivative,
we have

\begin{equation}
m^{\mu}_i =
{1 \over K_E a(1-a)}
\sum_j c^{\kl EE}_{ij} (\xi^{\mu}_j-a)
F^\mu_{Ej}
\label{dm2}
+
{\beta \over K_E a(1-a)}
\sum_j c^{\kl EE}_{ij} (\xi^{\mu}_j-a) \xi_j^\mu m^\mu_j
F^{\mu'}_{Ej} 
\, .
\end{equation}

\bigskip

\noindent
We can write Eq.\ (\ref{dm2}) in matrix form as 
\begin{equation}
({\mathbf I}-{\mathbf \Lambda}^\mu){\mathbf m^{\mu}}={\mathbf \Xi}^\mu
\, ,
\label{sel_cont_mat}
\end{equation}

\noindent
where ${\mathbf I}$ is the identity matrix, the $i^{\rm th}$
component of ${\mathbf m^{\mu}}$ is equal to $m^{\mu}_i$, and
the matrices ${\mathbf \Lambda}^\mu$ and ${\mathbf \Xi}^\mu$ are given by

\begin{subequations}
\begin{align}
\Lambda_{ij}^\mu & = \frac{\beta}{K_E a(1-a)}
\, c^{EE}_{ij}
(\xi^{\mu}_j-a)\xi^{\mu}_j
F^{\mu'}_{Ej}
\label{matrices.a}
\\
\Xi_{ij}^\mu & = \frac{1}{K_E a(1-a)}
\, c^{EE}_{ij}
(\xi^{\mu}_j-a) F^\mu_{Ej}
\, .
\label{matrices.b}
\end{align}
\label{matrices}
\end{subequations}

To solve Eq.\ (\ref{sel_cont_mat}) we need to invert
${\mathbf I} - {\mathbf \Lambda}$, in general a hard problem. However,
what we show now is that ${\mathbf \Lambda}$ has only one $\order(1)$ eigenvalue, with the rest
$\order(K_E^{-1/2})$. This allows us to write the inverse in terms of the
a single eigenvector and adjoint eigenvector, a simplification that
allows us to perform the inversion explicitly.

The spectrum of the random matrix, $\b \Lambda^\mu$,
is determined primarily by the mean
and variance of its components \cite{Mehta91}. In the large $N_E$ limit, 
these are given by

\begin{subequations}
\begin{align}
\left \langle \Lambda^{\mu}_{ij} \right \rangle_{ij} &=
\frac{\beta}{N_E} \left\langle F'_\k{E}(h_\k E+\beta \xi m+\hat{\sigma}_\k E z) \right\rangle_{\xi,z}\nonumber
\\
\left \langle {\Lambda^{\mu}_{ij}}^2 \right \rangle_{ij}-\left \langle \Lambda^{\mu}_{ij} \right \rangle^2_{ij}
&= \frac{\beta^2} {a K_E N_E} \left\langle {F'^2_\k {E}(h_\k E+\beta \xi m+\hat{\sigma}_\k E z)} \right\rangle_{\xi,z} \,\nonumber
\end{align}
\end{subequations}

\noindent
where $\left \langle \cdots \right \rangle_{ij}$ indicates an
average over $i$ and $j$, and 
we used the fact that $\xi_j^\mu$ and $F^{\mu'}_{Ej}$ are independent.

Given that ${\mathbf \Lambda}^\mu$ is an $N_E\times N_E$ matrix,
the fact that the mean and variance of its elements are $\order(N_E^{-1})$
and $\order ((K_E N_E)^{-1})$, respectively, implies that it 
has one eigenvalue that is $\order(1)$
and $N_E-1$ eigenvalues that are $\order(K_E^{-1/2})$ \cite{Mehta91}.
Letting $\b v_k$ and $\b v^\dagger_k$ be the eigenvector and adjoint
eigenvector of $\b \Lambda^\mu$ whose
eigenvalue is $\lambda_k$,
we can solve Eq.\ (\ref{sel_cont_mat}) for $\b m^\mu$,
\begin{equation}
\b m^{\mu} =
\sum_k {\b v_k \b v_k^{\dagger} \cdot {\mathbf \Xi}^\mu \over 1-\lambda_k}\nonumber
\end{equation}

\noindent
where ``$\cdot$'' represents dot product. Letting $k=0$ correspond to
the $\order(1)$ eigenvalue and explicitly
separating out this component, the expression for $\b m^\mu$ becomes

\begin{eqnarray}
{\mathbf m^{\mu}} &&=
\sum_{k\neq 0}
{\b v_k \b v_k^{\dagger} \cdot {\mathbf \Xi}^\mu \over 1-\lambda_k}
+
{\b v_0 \b v_0^{\dagger} \cdot {\mathbf \Xi}^\mu \over 1-\lambda_0}
\nonumber
\\
&&\approx
\sum_{k\neq 0} \b v_k \b v_k^{\dagger} \cdot {\mathbf \Xi}^\mu
+
{\b v_0 \b v_0^{\dagger} \cdot {\mathbf \Xi}^\mu \over 1-\lambda_0}
\label{matrix_sol}
\\
&&=
\sum_k \b v_k \b v_k^{\dagger} \cdot {\mathbf \Xi}^\mu
- \b v_0 \b v_0^{\dagger} \cdot {\mathbf \Xi}^\mu
+
{\b v_0 \b v_0^{\dagger} \cdot {\mathbf \Xi}^\mu \over 1-\lambda_0}
\nonumber
\\
&&=
\b \Xi^\mu
+ {\lambda_0 \over 1-\lambda_0} \,
\b v_0 \b v_0^{\dagger} \cdot {\mathbf \Xi}^\mu
\, 
\nonumber
\end{eqnarray}
and 
\begin{equation}
\lambda_0=\beta \langle F'_\k {E}(h_\k E+
\beta \xi m+\hat{\sigma}_E z) \rangle_{\xi,z}.
\label{lambda_zero}
\end{equation}

\noindent
Since $\b v_0$ and $\b v_0^\dagger$ are vectors whose components
are all the same, without loss of generality we can choose
$\b v_0 = (1, 1, ..., 1)/N_E$ and $\b v_0^\dagger = (1, 1, ..., 1)$.
Combining this choice with Eq.\ (\ref{matrix_sol})
and using Eq.\ (\ref{matrices.b}) for $\b \Xi^\mu$, we have
\begin{equation}
m^{\mu}_i=\sum_j\left[c^{EE}_{ij}+\frac{c\lambda_0}{1-\lambda_0}\right]
\frac{(\xi^{\mu}_j-a)F^\mu_{Ej}}{K_Ea(1-a)}.
\label{solution_m_nu1}
\end{equation}

We are now in a position to return to Eq.\ (\ref{var1}) and compute
the variance of $\delta h_m$ (which, recall, is denoted $\rho^2$).
Treating, as usual, all the terms in Eq.\ (\ref{var1}) as independent,
we have
\begin{equation}
\rho^2 =
\beta^2
\sum_{\mu \nu}
\langle \xi_i^\mu  \xi_i^{\nu} \rangle_{\xi}
\langle m^{\mu}_i m^{\nu}_i \rangle_{\xi,z}
=
p\beta^2 a
\langle m^{\mu^2}_i \rangle_{\xi,z}
\, .
\label{rho1}
\end{equation}

\noindent
To compute $\langle m^{\mu 2}_i \rangle_{\xi,i}$ we use
Eq.\ (\ref{solution_m_nu1}) and the fact that the 
off-diagonal elements average to zero, and we find that

\begin{equation}
\langle m^{\mu 2}_i \rangle_{\xi,z}
=
N_E \left[
c + 
{2 c^2 \lambda_0 \over (1-\lambda_0)} +
{c^2 \lambda_0^2 \over (1-\lambda_0)^2}
\right]
{\langle F^2_\k E(h_\k E+\beta \xi m+\hat{\sigma}_\k E z) \rangle_{\xi,z} \over K_E^2 a(1-a)}
\, .
\label{var_delta_mu}
\end{equation}

\bigskip

\noindent
To derive this expression we again used we used $\langle (\xi - a)^2 \rangle = a(1-a)$.

Our final step is to insert Eq.\ (\ref{var_delta_mu})
into (\ref{rho1}).
Ignoring the two terms in brackets in
Eq.\ (\ref{var_delta_mu}) that are a factor of $c$
smaller than
the first, and using the fact that
$\langle F^2_\k E(h_\k E+\beta \xi m+\hat{\sigma}_\k E z) \rangle_{\xi,z} = \langle \nu_\k E^2 \rangle$,
this leads to the
expression for $\rho^2$ given in Eq.\ (\ref{var_m}).
Consequently, loop corrections vanish, and we can use our naive
estimate for the variance of $\delta h_m$.

Ignoring the two terms in brackets in Eq.\ (\ref{var_delta_mu}) is
strictly correct  for infinitely diluted networks; i.e., networks with
$c \rightarrow 0$. When $c$ is nonzero but small, the terms in the
brackets can be ignored safely unless $\lambda_0\rightarrow 1$.
However, as we now show, $\lambda_0 = 1$ is precisely the point where
the background becomes unstable. Thus, it is not a regime in which we
can operate.

The significance of the limit $\lambda_0 = 1$ can be seen by replacing
Eq.\ (\ref{sel_cont_mat}) by it's dynamical counterpart
(see Eq.\ (\ref{dynamic_fr})),
\begin{equation}
\tau_E \, {d \b m^\mu \over dt} =
({\mathbf \Lambda}^\mu - {\mathbf I}){\mathbf m^{\mu}} + {\mathbf \Xi}^\mu
\, .
\end{equation}

\noindent
When the largest eigenvalue of $\Lambda^\mu$ exceeds 1, the
unactivated memories become unstable, and retrieval of just one memory
is impossible. As discussed above, the largest eigenvalue of
$\Lambda^\mu$ is $\lambda_0$. Consequently, loop corrections are
necessarily important (no matter how dilute the network is) at
precisely the point where the unactivated memories,
and thus the background, become unstable.

\renewcommand{\theequation}{III-\arabic{equation}}
\renewcommand{\thefigure}{III-\arabic{figure}}
\setcounter{equation}{0}
\setcounter{figure}{0}
\section*{Appendix III: Stability analysis}

To determine stability, we need to write down time-evolution equations
for the order parameters, and then linearize those around their fixed
points. For $\nu_\k E$, $\nu_\k I$ and $m$, which
are linear combinations of the firing rates, this is straightforward
-- we simply insert their definitions, Eq.\ (\ref{mf_sum}), into the
time-evolution equations for the individual firing rates,
Eq.\ (\ref{dynamic_fr}). For the variances,
$\hat{\sigma}_E^2$ and $\sigma_I^2$, the situation is much more
difficult, as these quantities do not admit simple time-evolution
equations \cite{Cool01}. Fortunately, we expect the
effects of the variances to be small -- as discussed in the main text,
their primary effect is to smooth slightly the gain functions, something
that typically (although presumably not always) stabilizes the
dynamics. Alternatively, if we assume that the variances are functions
of $\nu_\k E$, $\nu_\k I$ and $m$ (meaning we give them
instantaneous dynamics), we can rigorously neglect them.
This is because derivatives of the gain functions with respect to
$\nu_\k E$ and $\nu_\k I$ are large, on the order of $K^{1/2}$, while
derivatives with respect to the variances are $\order(1)$.
Thus, as a first approximation, we will ignore these
variables, and consider only the dynamics of
$\nu_\k E$, $\nu_\k I$ and $m$. Because
of this approximation, we expect our stability boundaries to be off by
a small amount.

Combining Eqs.\ (\ref{dynamic_fr}) and (\ref{MF}), the time-evolution
equations for $\nu_\k E$, $\nu_\k I$ and $m$ may be written

\begin{subequations}
\begin{align}
\tau_\k E \, {d \nu_\k E \over dt} &= \left\langle
\langle F_E(h_E + \xi \beta m + \hat{\sigma}_E z) \rangle_{\xi}
\right\rangle_z
- \nu_\k E
\label{MF_t.a}
\\
\tau_\k I \, {d \nu_\k I \over dt} &=
\left\langle F_I(h_I + \sigma_I z) \right\rangle_z
- \nu_\k I
\label{MF_t.b}
\\
\tau_\k E \, {d m \over dt} &= \left\langle
\langle (a(1-a))^{-1} (\xi-a)
F_E(h_E + \xi \beta m + \hat{\sigma}_E z) \rangle_{\xi}
\right\rangle_z
- m
\label{MF_t.c}
\, .
\end{align}
\label{MF_t}
\end{subequations}

\noindent
To simplify notation, it is convenient to define

\begin{subequations}
\begin{align}
\phi_E(\nu_\k E, \nu_\k I, m) & \equiv \left\langle
\langle F_E(h_E + \xi \beta m + \hat{\sigma}_E z) \rangle_{\xi}
\right\rangle_z
\label{MF_defs.a}
\\
\phi_I(\nu_\k E, \nu_\k I, m) & \equiv
\left\langle F_I(h_I + \sigma_I z) \right\rangle_z
\label{MF_defs.b}
\\
\phi_m(\nu_\k E, \nu_\k I, m) & \equiv \left\langle
\langle (a(1-a))^{-1} (\xi-a)
F_E(h_E + \xi \beta m + \hat{\sigma}_E z) \rangle_{\xi}
\right\rangle_z
\label{MF_defs.c}
\, .
\end{align}
\label{MF_defs}
\end{subequations}

\noindent
Then, linearizing Eq.\ (\ref{MF_t}) by letting
$\nu_\k E \rightarrow \nu_\k E + \delta \nu_\k E$,
$\nu_\k I \rightarrow \nu_\k I + \delta \nu_\k I$,
and
$m \rightarrow m + \delta m$, we have

\begin{equation}
{d \over dt}
\left( \begin{array}{c}
\delta \nu_\k E
\\
\delta \nu_\k I
\\
\delta m
\end{array} \right)
=
\left( \begin{array}{ccc}
\tau_E^{-1} (\phi_{E,E} - 1) &
\tau_E^{-1}  \phi_{E,I} &
\tau_E^{-1}  \phi_{E,m}
\\
\tau_I^{-1}  \phi_{I,E} &
\tau_I^{-1} (\phi_{I,I} - 1) &
\tau_I^{-1}  \phi_{I,m}
\\
\tau_E^{-1}  \phi_{m,E} &
\tau_E^{-1}  \phi_{m,I} &
\tau_E^{-1} (\phi_{m,m} - 1)
\end{array} \right)
\left( \begin{array}{c}
\delta \nu_\k E
\\
\delta \nu_\k I
\\
\delta m
\end{array} \right)
\nonumber 
\end{equation}

\bigskip

\noindent
where the notation $\phi_{a,b}$ indicates a derivative of $\phi_a$
with respect to the argument specified by $b$ (for example,
$\phi_{E,I} = \partial \phi_E / \partial \nu_\k I$ and
$\phi_{I,m} = \partial \phi_I / \partial m$). Since $\phi_I$ is
independent of $m$ (which means $\phi_{I,m} = 0$), the equation for the
eigenvalues, denoted $\lambda$, becomes
\begin{eqnarray}
\nonumber
0 & = & [((\phi_{E,E} - 1) - \tau_E \lambda)
((\phi_{I,I} - 1) - \tau_I \lambda) -
\phi_{E,I} \, \phi_{I,E}]
((\phi_{m,m} - 1) - \tau_E \lambda)
\\
& + &
[\phi_{I,E} \,  \phi_{m,I} -
((\phi_{I,I} - 1) - \tau_I \lambda) \phi_{m,E}]
\phi_{E,m}
\, .
\label{eigen}
\end{eqnarray}

Equation (\ref{eigen}) is a cubic equation in $\lambda$, and thus not
straightforward to solve. However, in the large $K$ limit it
simplifies considerably. That's because derivatives with respect to
$\nu_\k E$ and $\nu_\k I$ are $\order(K^{1/2})$, which follows because
the $\phi$'s depend on $\nu_\k E$ and $\nu_\k I$ through $h_\k E$ and
$h_\k I$, and the latter are proportional to $K^{1/2}$
(see Eq.\ (\ref{he_hi})). Defining the $\order(1)$ quantities
\begin{equation}
\phi_{a,Q}^0 \equiv
K^{-1/2} \phi_{a,Q}\nonumber
\, ,
\end{equation}

\noindent
$a= \nu_\k E, \nu_\k I, m$ and $Q = \nu_\k E, \nu_\k I$,
Eq.\ (\ref{eigen}) becomes (ignoring $\order(K^{-1/2})$ corrections)
\begin{eqnarray}
\nonumber
0 & = & [(\phi_{E,E}^0 - K^{-1/2} \tau_E \lambda)
(\phi_{I,I}^0 - K^{-1/2} \tau_I \lambda) -
\phi_{E,I}^0 \, \phi_{I,E}^0]
((\phi_{m,m} - 1) - \tau_E \lambda)
\\
& + &
[\phi_{I,E}^0 \,  \phi_{m,I}^0 -
(\phi_{I,I}^0 - K^{-1/2} \tau_I \lambda) \phi_{m,E}^0]
\phi_{E,m}
\, .
\label{eigen0}
\end{eqnarray}

Examining Eq.\ (\ref{eigen0}), it follows that if the eigenvalue,
$\lambda$, is $\order(K^{1/2})$, then
the term $\phi_{m,m}-1$ and the last term in brackets can
be neglected. There
are two such eigenvalues, and they are given by
\begin{equation}
\lambda_{\pm} = {
\tau_E^{-1} \phi_{E,E} + \tau_I^{-1} \phi_{I,I}
\pm \{
(\tau_E^{-1} \phi_{E,E} + \tau_I^{-1} \phi_{I,I})^2
- 4 [(\tau_E \tau_I)^{-1}(\phi_{E,E} \phi_{I,I} - \phi_{E,I} \phi_{I,E}]
\}^{1/2}
\over 2}\nonumber
\, .
\end{equation}

\noindent
Both eigenvalues are negative if
\begin{subequations}
\begin{align}
\tau_E^{-1} \phi_{E,E} + \tau_I^{-1} \phi_{I,I} & < 0
\\
\phi_{E,E} \phi_{I,I} - \phi_{E,I} \phi_{I,E} & > 0
\, .
\label{conditions.a}
\end{align}
\label{conditions}
\end{subequations}

\noindent
Since $\phi_{I,I} < 0$, the first condition is satisfied if $\tau_I$
is sufficiently small. For the second condition, from
Eqs.\ (\ref{he_hi}), (\ref{detJ}) and (\ref{MF_defs}) we see that
\begin{equation}
\phi_{E,E} \phi_{I,I} - \phi_{E,I} \phi_{I,E} \propto D \nonumber
\end{equation}

\noindent
where the constant of proportionality is positive.
Since the condition for the stability of the background is $D > 0$
\cite{Van98}, we see that Eq.\ (\ref{conditions.a}) is
satisfied whenever the background is stable.
Thus, for $\tau_I$ sufficiently small and the background stable, the two
$\order(K^{1/2})$ eigenvalues are negative.

The third eigenvalue is
$\order(1)$, so when computing it we can drop all
the $K^{-1/2} \lambda$ terms. Denoting this eigenvalue $\lambda_m$, we thus
have

\begin{equation}
\lambda_m = \phi_{m,m} - 1 +
{(\phi_{I,E} \phi_{m,I} - \phi_{I,I} \phi_{m,E}) \, \phi_{E,m} \over
\phi_{E,E} \phi_{I,I} - \phi_{E,I} \phi_{I,E}}
\, .
\label{lambda_small}
\end{equation}

\noindent
Using a prime to denote a derivative with respect to $h_E$ and noting that (see Eqs.\ (\ref{MF_defs}))
\begin{eqnarray}
&&\phi_{Q,R}=\frac{\partial \phi_Q}{\partial h_Q} \frac{\partial h_Q}{\partial \nu_R}\nonumber\\
&&\phi_{m,R}=\frac{\partial \phi_m}{\partial h_E} \frac{\partial h_E}{\partial \nu_R}\nonumber,
\end{eqnarray}

\noindent Eq.\ \ref{lambda_small} reduces to
\begin{equation}
\lambda_m = \phi_{m,m} - 1 - {\phi'_m \phi_{E,m} \over \phi'_E}\nonumber
\, ,
\end{equation}
where prime denotes a derivative. 

Comparing Eqs.\ (\ref{MF_defs}) and Eq.\ (\ref{MFx}), we see that $\phi_{E,m} = a \phi_{m,m}$,
which leads to
\begin{equation}
\lambda_m =
\phi_{m,m} \left( 1 - {a\phi'_m\over \phi'_E} \right) - 1
\, .
\label{lambda_small2}
\end{equation}

\noindent
This expression strongly emphasizes the role of the coding level, $a$:
if it were zero, the only stable equilibria would be those with
$\phi_{m,m} < 1$, which would imply high firing rates for foreground
neurons (see Fig.\ \ref{Fig1}b).

Although Eq.\ (\ref{lambda_small2}) tells us the stability of an 
equilibrium, it is not in an especially convenient form, as
it does not allow us to look at a set of nullclines and determine 
instantly which equilibria are stable and which are not.
However, it turns out that it is rather easy to 
determine the sign of $\lambda_m$ for a given set of nullclines
simply by looking at them. To see how,
we make use of the expressions
for $\phi_E$ and $\phi_m$ (Eqs.\ (\ref{MF_defs.a}) and (\ref{MF_defs.c})) to 
reduce the right hand side of Eq.\ (\ref{lambda_small2}) to an expression with 
a single derivative. Our starting point is the definition 
\begin{equation}
\Psi(m)\equiv \bar{F}_E( h_\k E(m) + \beta m) - \bar{F}_E( h_\k E(m)),
\label{Psi}
\end{equation}
where $h_E(m)$ is given by Eq.\ (\ref{hem}); the solutions of the 
equation $\Psi(m)=m$ correspond to network equilibria.
The advantage of this one dimensional formulation is that, as we 
show below, the condition $\lambda_m<0$ is 
equivalent to $d\Psi/dm <1$. Thus, 
by plotting the function $\Psi(m)$ versus $m$ and looking at its 
intersections with the $45^{\circ}$ line, we can find the 
equilibrium values of $m$, and, more importantly, we can easily determine
which of them is stable and which is unstable. 

To show that $d\Psi(m)/dm< 1$ is equivalent to the condition $\lambda_m<0$,
we note first of all that 

\begin{eqnarray}
&&\phi_{m,m}=\beta \bar{F}'_E(h_\k E+ \beta m)\nonumber\\
&&\phi'_m=\bar{F}'_E(h_\k E+ \beta m)-\bar{F}'_E(h_\k E) \nonumber\\
&&\phi'_E=a \bar{F}'_E(h_\k E+ \beta m)+ (1-a) \bar{F}'_E(h_\k E),\nonumber
\end{eqnarray}

\noindent where, recall, a prime denotes a derivative. By combining these expressions with  Eq.\ (\ref{lambda_small2}),
and performing a small amount of algebra, the condition $\lambda_m<0$ can be written
\begin{equation}
\beta \bar{F}'(h_\k E+ \beta m) \bar{F}'(h_\k E) < a
\bar{F}'(h_\k E+ \beta m)+ (1-a) \bar{F}'(h_\k E).
\label{lambdag1}
\end{equation}

\noindent To see how this compares to $d\Psi/dm$, we use Eq.\ (\ref{Psi})
to write

\begin{equation}
\frac{d \Psi}{dm}= \beta \bar{F}'_E(h_\k E+ \beta m) +
[\bar{F}'_E(h_\k E+ \beta m)- \bar{F}'_E(h_\k E)] \frac{dh_\k E}{dm}.
\nonumber
\end{equation}

\bigskip

\noindent Then, using Eq.\ (\ref{he_nullcline}), which tells us that 
\begin{equation}
\frac{dh_\k E}{dm}=\frac{-a}{\bar{F}'_E(h_\k E)},\nonumber
\end{equation}

\noindent this expression becomes

\begin{equation}
\frac{d\Psi}{dm}=1+\frac{\beta \bar{F}'_E(h_\k E+ \beta m)\bar{F}'_E(h_\k E)-[a\bar{F}'_E(h_\k E+ \beta m)+(1-a)\bar{F}'_E(h_\k E)]}{a\bar{F}'_E(h_\k E)}.
\label{psiprmg1}
\end{equation}

\bigskip

\begin{figure}
\centering
\subfigure{\includegraphics[height=6.5cm,width=7cm]{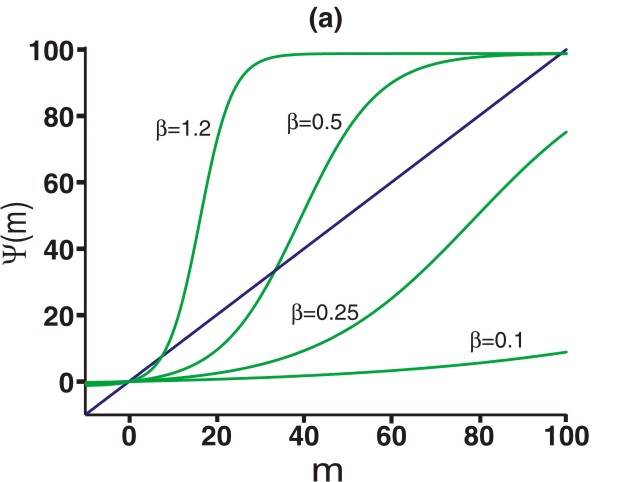}}
\subfigure{\includegraphics[height=6.5cm,width=7cm]{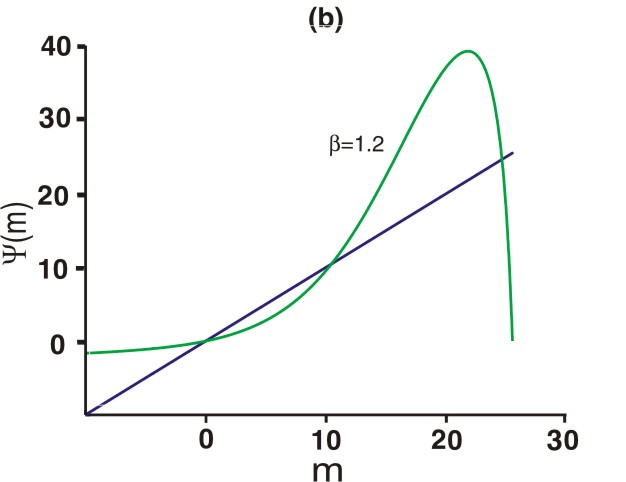}}
\caption{{\bf (a)} $\Psi(m)$ versus $m$ for $a=0.001$ and 
$\beta=0.1, 0.25,0.5$ and $1.2$. The
equilibrium values of $m$ are the intersections 
of the curves $\Psi(m)$ with the diagonal line.
For small $\beta$, $m=0$ is the only solution. 
For intermediate $\beta$ there are two additional non-zero solutions.
{\bf (b)} $\Psi(m)$ versus $m$ for $a=0.05$ and $\beta=1.2$; the upper intersection
is now at a biologically realistic firing rate. Note different scales compared to (a). 
Other parameters, as in Fig.\ 2, are $J_\k {EE}=J_\k {IE}=1, J_\k {EI}=-1.9, J_\k {II}=-1.5, h_\k {Eex}=3, h_\k {Iex}=2.1$.}
\label{Fig3}
\end{figure}

\noindent
Comparing Eqs.\ (\ref{lambdag1}) and (\ref{psiprmg1}), we see that
the condition $d\Psi/dm < 1$ is equivalent to $\lambda_m < 0$. Thus, it
is only when $\Psi(m)$ intersects the $45^{\circ}$ line from above
that the equilibrium is stable. Since $\Psi(m)$ is bounded, if there
are three equilibria, the smallest one must be stable, the middle one unstable 
and the largest one again stable. Thus, we can look at the nullcline plots and 
immediately determine stability (see below and Fig. \ref{Fig3}). 

As an example we revisit Fig.\ \ref{Fig2}.
In terms of our specific form for the gain functions, 
Eq.\ (\ref{tr_func}), and with $h_\k E(m)$ given by Eq.\ (\ref{hem}), 
the equation for $m$ becomes

\begin{eqnarray}
m=\Psi(m)=
\nu_{\max}
H\left[
H^{-1} \left( {\nu_{\k E0}-am \over \nu_{\max}} \right) +
{\beta m \over \sigma_E}
\right]-(\nu_{\k E0}-am)
\, .
\label{self-const}
\end{eqnarray}

\noindent This equation is solved graphically in Fig.\ \ref{Fig3}a where we plot
$\Psi(m)$ versus $m$ for the same values of $\beta$ used in Fig.\ \ref{Fig2} 
and with $a=0.005$. Intersections with the $45^\circ$ line correspond
to solutions of Eq.\ (\ref{self-const}), and thus to network
equilibria. 

As we saw in sections \ref{red-MF} and \ref{exm-binary}, the main factor that determines the
number and location of the intersections, and thus the ability of the
network to exhibit retrieval states, is $\beta$.
For $\beta=0.1$ and $0.25$, there is just one intersection at
$m=0$, while for intermediate values of $\beta$, $\beta=0.5$ and $1.2$,
two additional intersections appear. Increasing $\beta$ 
even further moves one of the solutions to negative $m$
and destabilizes the background, but this is not shown. We can now 
easily see that the curves in
Fig.\ \ref{Fig3}a with $\beta = 0.1$ and $0.25$ have a single 
stable intersection at $m=0$ (meaning that the solutions with $m=0$ 
in Figs.\ \ref{Fig2}a and b are stable); 
the curves with $\beta=0.5$ and $\beta=1.2$ 
have two stable intersections, one at $m=0$ and one at large $m$ 
(and thus the solutions at $m=0$ in Fig.\ \ref{Fig2}c are stable, 
those at intermediate $m$ are unstable, and those with large $m$ 
are again stable). 

Although we see bistability, the firing rate for 
the retrieval state is unrealistically high -- on the order of 
$100$ Hz, near saturation. As discussed in the main text, 
we can reduce the firing rate by increasing $a$. This is done in 
Fig.\ \ref{Fig3}b, where we plot $\Psi(m)$ versus $m$ but this 
time for $a=0.05$ and $\beta=1.2$. Again there are three intersections 
(corresponding to the three intersections between the $m$-nullcline 
with $\beta=1.2$ and the $h_E$-nullcline with $a=0.05$ in 
Fig.\ \ref{Fig2}c). With this higher value of $a$, the upper 
intersection is now in a biologically realistic range.

\renewcommand{\theequation}{IV-\arabic{equation}}
\renewcommand{\thefigure}{IV-\arabic{figure}}
\setcounter{equation}{0}
\setcounter{figure}{0}
\section*{Appendix IV: Retrieval states in the finite connectivity regime}

When we performed network simulations, we found that the memory
strength, $\tilde{\beta}$, did not exhibit exactly the predicted $1/K$
scaling. Here we ask whether the departure from predictions that we
observed can be explained by finite $K$ corrections. These
corrections, as we will see shortly, are on the order of $K^{-1/2}
$. Since in our simulations $K$ is as small as 1,500,
these corrections are
potentially large.

Our starting point is the exact set of reduced mean field equations,
which is found by combining
Eqs.\ (\ref{balanced_cond}) and (\ref{nu0}),

\begin{subequations}
\begin{align}
\nu_{\k E0} +
K^{-1/2} D^{-1} \left[ J_{II} h_E - J_{EI} h_I \right]
&= \bar{F}_E( h_E) + am \nonumber\\
m &=
\bar{F}_E( h_E + \beta m) - \bar{F}_E( h_E) \nonumber\\
\nu_{\k I0} +
K^{-1/2} D^{-1} \left[ J_{EE} h_I - J_{IE} h_E \right]
&= \bar{F}_I( h_I)\, . \nonumber
\end{align}
\end{subequations}

\noindent
When $K$ is large we can solve these equations by perturbing around
the $K \rightarrow \infty$ solutions, which we denote $h_{E0}$, $m_0$
and $h_{I0}$ (these are the solutions to Eq.\ (\ref{mf_tilde0})).
The zeroth step in this perturbation analysis is to replace $h_E$ and
$h_I$ by $h_{E0}$ and $h_{I0}$ where they appear in brackets (and thus
multiply $K^{-1/2}$). This gives us a new set of equations,

\begin{subequations}
\begin{align}
\nu_{\k E0} + \delta \nu_\k E
&= \bar{F}_E( h_E) + am
\label{dmf_tilde_hEk}\\
m &=
\bar{F}_E( h_E + \beta m) - \bar{F}_E( h_E)
\label{dmf_tilde_mk}
\\
\nu_{\k I0} + \delta \nu_\k I
&= \bar{F}_I( h_I)
\label{dmf_tilde_hIk}
\end{align}
\label{dmf_tildek}
\end{subequations}

\noindent
where

\begin{subequations}
\begin{align}
\delta \nu_E & \equiv
K^{-1/2} D^{-1} \left[ J_{II} h_{E0} - J_{EI} h_{I0} \right]
\\
\delta \nu_E & \equiv
K^{-1/2} D^{-1} \left[ J_{EE} h_{I0} - J_{IE} h_{E0} \right]
\, .
\end{align}
\label{dnu}
\end{subequations}

For the inhibitory firing rate, it is easy to see the effect of finite
$K$: $h_I$ is shifted relative to $h_{I0}$ by an amount proportional
to $\delta \nu_\k I$. Only slightly more difficult are $h_E$ and $m$,
for which we have to consider how $\delta \nu_\k E$ affects the
nullclines. Fortunately, only the $h_E$-nullcline is affected, and we
see that it shifts in a direction given by the sign of $\delta \nu_\k
E$. In particular,
\begin{equation}
{d (-h_E) \over d \delta \nu_\k E} = {-1 \over \bar{F}'(h_E)}
\, .
\label{dhEdnu}
\end{equation}

\bigskip

\noindent
(We consider -$h_E$ since,
by convention we plot our nullclines in a space with
-$h_E$ on the $y$-axis.) Thus, if
$\delta \nu_\k E$ is positive then the $h_E$-nullcline shifts
down relative to $h_{E0}$,
while if it is negative the nullcline shifts up.

In our simulations we set $\beta$ to $\beta_{min}$, the minimum
value of $\beta$ that allows retrieval of one memory. To
determine how $K$ affects $\beta_{min}$, then, we need to know how to
adjust $\beta$ so that we keep the grazing intersection
as $K$ changes. Fortunately, the
$h_E$-nullcline depends on $K$ but not $\beta$, and the $m$-nullcline
depends on $\beta$ but not $K$. Thus, all we need to know is how the
$m$-nullcline changes with $\beta$. Using Eq.\
(\ref{dmf_tilde_mk}), it is easy to show that at fixed $m$,

\begin{equation}
{d (-h_E) \over d \beta} =
{m \bar{F}'(h_E + \beta m) \over
\bar{F}'(h_E + \beta m) - \bar{F}'(h_E)}
\, .
\label{dhEdbeta}
\end{equation}

\bigskip

\noindent
The numerator in this expression is clearly positive and,
for equilibria to the left of the peak of the $m$-nullcline, the
denominator is also positive (see Appendix III). Thus, increasing
$\beta$ causes the $m$-nullcline to move up.

\begin{figure}
\centering
\includegraphics[height=6.5cm,width=7cm]{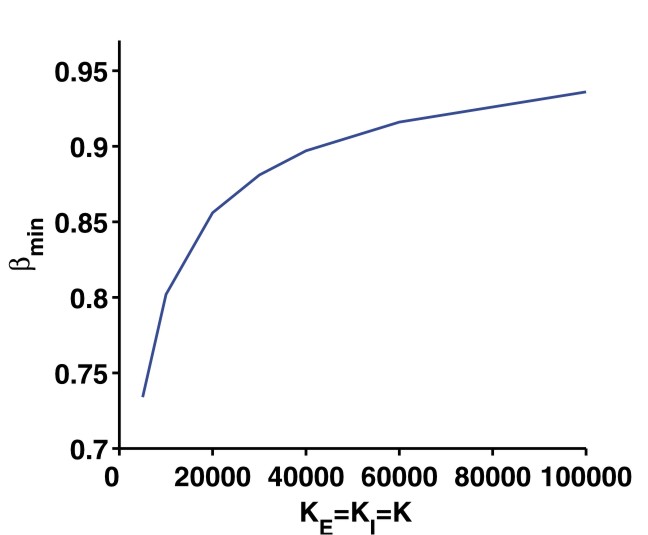}
\caption{The effect of finite $K$ corrections on $\beta_{min}$. 
The minimum value of $\beta$ at which a single stored 
pattern can be retrieved successfully, $\beta_{min}$, decreases as $K$ decreases. The parameters are the same as 
Fig.\ 2 with $a=0.05$.}
\label{Fig5}
\end{figure}

Combining Eqs.\ (\ref{dhEdnu}) and (\ref{dhEdbeta}),
we have the following picture,

\[
\delta \nu_\k E      \ {\rm decreases}
\implies h_E{\rm -nullcline\ moves\ up}
\implies \beta_{min} \ {\rm increases}
\, 
\]

\noindent
where ``up'' corresponds to movement in the $m-(-h_\k E)$ plane.
To complete the picture, we need to know how $\delta \nu_\k E$
depends on $K$. From Eq.\ (\ref{dnu}), we see that
$\delta \nu_\k E \propto K^{-1/2} [J_{II} h_{E0} - J_{EI} h_{I0}]=
K^{-1/2} [-|J_{II}| h_{E0} + |J_{EI}| h_{I0}]$. Thus, whether
$\delta \nu_\k E$ is an increasing or decreasing function of $K$
depends on whether $|J_{II}| h_{E0}$ is larger or smaller
than $|J_{EI}| h_{I0}$. However, as we have seen, typically $h_E$ is
negative. Thus, we expect $\delta \nu_\k E$ to be proportional to
$K^{-1/2}$ with a positive constant of proportionality, which means
that $\delta \nu_\k E$ is a decreasing function of $K$. Combining that
with the above picture, we conclude that when $K$ increases
$\beta_{min}$ also increases. 
This is shown explicitly in Fig.\ \ref{Fig5}. Moreover, it was
exactly what we saw in our simulations: $\beta_{min}$ ($\tilde{\beta}$
in Table I) was larger than predicted when we increased $K$
(compare $\tilde{\beta}$ with $\tilde{\beta}_{predicted}$
in Table I).

\renewcommand{\theequation}{V-\arabic{equation}}
\renewcommand{\thefigure}{V-\arabic{figure}}
\setcounter{equation}{0}
\setcounter{figure}{0}
\section*{Appendix V: Fine tuning in the learning rule}
\label{fine_tuning}
In the model described here, the structured part of the synaptic
weights scale as $K^{-1}$ whereas the background scales as $K^{-1/2}$. This 
appears to require fine tuning, since adjustments to the weights during learning 
of the attractors have to be a factor of $K^{1/2}$ times smaller than the 
background weights; a factor that can be as high as 100. 

The first question to ask, then, is: exactly how big is the fine tuning problem?
In other words, how much noise can we add to the learning rule
without having a huge effect on the storage capacity? This can be answered by 
considering a learning rule in which the weight changes during learning a pattern
are not quite perfect. Specifically, let us consider the 
following modification of Eq.\ (\ref{A-ij}),

\begin{equation}
A_{ij}= \frac{\beta}{a(1-a)K_\k E} \sum_{\mu} \xi^{\mu}_i (\xi^{\mu}_j-a)+ \sum_{\mu} \eta^{\mu}_{ij}
\label{deltaA-ij},
\end{equation}

\noindent
where the $\eta^{\mu}_{ij}$ are zero-mean, uncorrelated random variables
with variance $\sigma_\eta^2$. The additional noise in this learning 
rule increases the variance of the quenched noise by an amount $K_\k E p \sigma^2_\eta$.
As a result, if 

\begin{equation}
\sigma^2_\eta\sim \order\left((K_\k E p)^{-1}\right), 
\label{noise-scaling}
\end{equation}

\noindent the effect on storage capacity is an $\order(1)$ increase in
the quenched noise, and thus the storage capacity still scales as $K_\k E$.

With the scaling in Eq. (\ref{noise-scaling}), weight changes during 
learning of each pattern is a factor of $p^{1/2}$ smaller than the background weights, and therefore 
the amount of fine tuning depends on how many patterns are stored. 
Because of the low storage capacity found in these networks (at most 2.5\% \cite{Lat04}), 
even when $K$ is as large as $10,000$, $p^{-1/2}$ is
on the order of 6\%. 

We should also point out that it is possible for the weight
changes associated with the structured part of the connectivity to be on the
same order as the background, although at the expense of
storage capacity. Let us consider a third learning rule in which each synapse has
a probability $q$ of changing its value during learning,

\begin{equation} 
A_{ij}=\tilde{\beta}' \sum_{\mu} q^{\mu}_{ij}
\xi^{\mu}_{i}(\xi^{\mu}_{j}-a), 
\label{prob-learning}
\end{equation}

\noindent
where the $q^{\mu}_{ij}$ are Bernoulli variables;  $q^{\mu}_{ij}=1$ with
probability $q$ and $0$ with probability $1-q$. Let us define the coupling
strength slightly differently than in Eq.\ (\ref{beta-scaling}),
\begin{equation}
\tilde{\beta}' = \frac{\beta}{a(1-a)qK_\k E} \nonumber
\, ,
\end{equation}

\noindent
where, as usual, $\beta\sim \order(1)$. With this definition, the mean
memory strength, $\langle \tilde{\beta}' q_{ij}^\mu \rangle$, is again
$\beta/K_E a(1-a)$, as in Eq.\ (\ref{beta-scaling}).
But by setting $q\sim \order({K_\k E}^{-1/2})$, the synaptic weight change --
if there is one -- is $\order({K_\k E}^{-1/2})$, just as it is for the
background weights.
However, there is a major drawback: as is easy to show, the variance
associated with the structured part of the connectivity increases by a
factor of $K_\k E$, so the maximum number of patterns scales as
$p_{\max}\sim \sqrt{K_\k E}$ rather than $K_\k E$. We thus use
Eq.\ (\ref{A-ij}) for $A_{ij}$ in all of our analysis.
\bibliography{mybibliography}
\end{document}